\documentclass[aps,twocolumn,showpacs,prb,floatfix]{revtex4-1}
\usepackage[english]{babel}
\usepackage{amsmath}
\usepackage{amssymb} 
\usepackage{times}
\usepackage{graphicx}
\usepackage{color}
\usepackage{bbold}

\newcommand{\ave}[1]{\langle #1 \rangle}
\begin{document} 

\title{Charge Recombination in Undoped Cuprates }
         
\author{Zala Lenar\v ci\v c$^{1}$  and Peter Prelov\v sek$^{1,2}$}
\affiliation{$^1$J.\ Stefan Institute, SI-1000 Ljubljana, Slovenia}
\affiliation{$^2$Faculty of Mathematics and Physics, University of
Ljubljana, SI-1000 Ljubljana, Slovenia}

\begin{abstract}
We theoretically analyse the process of charge recombination in the planar Mott-Hubbard 
insulators with the aim to explain short picosecond-range lifetime of photoexcited carriers, experimentally
studied via pump-probe experiments on the undoped cuprates. The recombination mechanism consists of two essential ingredients: the formation of a metastable s-type bound holon-doublon pair,
i.e. the Mott exciton, and the decay of such an excitonic state via the multimagnon emission. In spite of 
the large gap that requires many bosons to be emitted, latter process is fast due to large exchange scale and strong 
charge-spin coupling in planar systems. As the starting microscopic model
we consider the single-band Hubbard model, and then more realistic three-band model for cuprates, both leading to the same minimal one. The decay rate of the exciton is 
evaluated numerically via the Fermi golden rule, having consistency also with the direct time-evolution
calculation. The decay rate reveals exponential dependence on the ratio of the Mott-Hubbard gap 
and the exchange coupling - the result qualitatively reproduced also within a toy exciton-boson 
model.

\end{abstract}


\maketitle

\section{Introduction}
Nonequilibrium properties and dynamics of strongly correlated electron systems are one of the central  
theoretical challenges, stimulated by the advances of ultrafast spectroscopy techniques and novel results 
in materials with correlated materials, as well as by the experiments on the fermionic cold atoms.  
One of the evident questions is the explanation of ultrafast recombination of  photoinduced charges, as established
in the pump-probe experiments on various materials belonging to the class of Mott-Hubbard (MH) insulators. The 
prominent example are undoped two-dimensional (2D) cuprates La$_2$CuO$_4$ (LCO) and 
Nd$_2$CuO$_4$ (NCO) , representing
the reference substances for the hole-doped and electron-doped high-$T_c$ superconductors, respectively.
The femtosecond pump-probe spectroscopy \cite{matsuda94,okamoto10,okamoto11} reveals that pump pulses
with photon energies above the MH gap $\Delta$ generate mobile charges, recombining in the picosecond range. 
This scale is 
many orders of magnitude shorter than in clean band insulators and semiconductors with similar gaps \cite{yu99}.
Photoexcited carriers in the MH insulators are in comparison to bosonic spin and phonon excitations a high-energy excitation far from equilibrium. Therefore the charge recombination process in a clean system requires an 
instantaneous emission  of the energy $\omega > \Delta$, which demands creation of many low-energy excitations, 
limiting the decay rate. The evident low-energy candidates in 2D cuprates are spin excitations with the characteristic spin exchange scale $J$, since as the consequence of strong correlations the effective charge-spin coupling is inherently strong, and also larger than the characteristic phonon energies $\omega_0$.
Similar questions extend to other MH materials, e.g. to the class of 
one-dimensional (1D) organic insulators where an ultrafast decay of photoinduced carriers was observed as well \cite{iwai03,uemura08,miyagoe08,mitrano14}.
Closely related is the challenge of fermionic cold atoms in optical lattices where near the
half-filled case the double-occupancy decay is somewhat faster \cite{strohmaier10,sensarma10}, yet still requires many scattering processes.

Theoretical analysis of strongly correlated electron systems far from equilibrium requires novel concepts 
and methods due to the failure of quasiparticle picture and Boltzmann-type approaches standard for 
metals and semiconductors. The relevant charge excitations in MH insulators, at least within the simplest 
prototype single-band Hubbard model, are empty sites - holons and doubly occupied sites - doublons.
At low holon-doublon densities latter excitations bear some resemblance to the holes and electron quasiparticles 
in semiconductors: a) they are oppositely charged relative to the reference insulator, b) they are well mobile with an 
effective band dispersion within the lower and upper Hubbard band, respectively, and c) they can form a bound 
excitonic-like state, i.e. a holon-doublon (HD) exciton. On the other hand, unlike in a pure semiconductor a 
single HD pair (neglecting the coupling to phonon degrees of freedom) is not an eigenstate and has an 
intrinsic recombination rate $\Gamma$. 

The problem of doublon decay  has been addressed in the Fermi-Hubbard model in connection with ultracold 
fermions in optical lattices \cite{sensarma10,strohmaier10} using the diagrammatic approach revealing an 
exponential dependence of the decay rate on the MH onsite repulsion $U$. 
Since in the latter case charge densities are quite high, the dominant mechanism relies on energy transfer to the kinetic energy of other fermions.
The decay of double occupancy was considered also within the excited half-filled
Hubbard model via the  time-dependent single-site dynamical mean-field theory (DMFT) 
\cite{eckstein11,eckstein12}, for review see \cite{aoki14}, confirming similar $\Gamma(U)$ dependence  that suggests the same recombination mechanism.
One should note that besides being at rather high effective temperatures $T$, by construction the DMFT method does not incorporate non-local spin fluctuations. Recombination of HD pair into spin excitations at low and high temperatures has already been addressed within the $n$th order perturbation theory\cite{sensarma11}. However, possible correlations between holon and doublon, i.e. the HD binding (an essential ingredient of our work) were neglected, since the prime interest was actually the decay of unpaired fermions in attractive Hubbard model.

Considering the case of finite photoexcited HD-pair densities $n_{HD}>0$ the recombination
processes could be qualitatively classified in 
analogy to semiconductors via the density dependence of recombination rates $\Gamma$, $\Gamma \propto n_{HD}^\gamma$,
into a single exponential one with $\gamma=0$,  bimolecular with $\gamma=1$ and Auger processes 
with $\gamma=2$.
We elaborate in this paper the charge-recombination scenario \cite{lenarcic13} relevant for undoped 
cuprates LCO and NCO, but also more generally for 2D MH insulators with a pronounced role of 
spin-fluctuation excitations. An important message from pump-probe experiments on those insulating cuprates 
\cite{okamoto10,okamoto11} is that is at least for modest pump fluences (pump intensity) the photoinduced charges (holons and doublons), measured via the probe broad-band optical pulse, decay exponentially after a very fast transient in the femtosecond range. 
The long-time decay rate in the picosecond range is fluence independent, i.e. independent on the
initial pump intensity and corresponding initial charge density. 
This excludes the 
interpretation in terms of bimolecular and Auger processes, and leaves the option with an 
intermediate stage of bound HD pairs - excitons, which decay exponentially with a well defined rate.
Relating back to the experiments, the initial fast transient should describe the relaxation of highly excited holons and doublons that end up in a bound HD exciton, but this is beyond our present study.
The existence of a bound MH exciton with the $s$ orbital symmetry has been shown within the planar Hubbard model that was for holons and doublons effectively reduced to 2D $t$-$J$ model \cite{tohyama06,lenarcic13}.
Due to its symmetry, exciton is not directly observable by optical absorption, but is
consistent with the experimental evidence of nonlinear optical susceptibility in LCO \cite{maeda04}, as well as 
a large Raman shift \cite{salamon95}. 

In a strongly correlated system the MH exciton is not an eigenstate of the system and 
can decay-recombine via the emission of spin fluctuations \cite{lenarcic13} with 
the characteristic boson scale $J$. Our first theoretical goal is to derive a 
proper perturbation term governing the decay. While in the initial study we start with the canonical transformation 
of the single-band Hubbard model,  \cite{lenarcic13}
undoped cuprates are known to be charge-transfer MH insulators. In the following we show that the effective 
HD recombination term emerging from a more complete multi-band model of cuprates
is even quantitatively similar to the one derived from the single-band model. 

The HD exciton decay with
the emission of a large number $n \sim \Delta/J \gg 1$ of spin fluctuations is an involved  many-body problem.
We calculate the recombination rate $\Gamma$ within the Fermi golden rule (FGR) approach, which still requires
a numerical evaluation  on a small-size system. Since our results are obtained on systems with limited size this implicitly shows that long-range antiferromagnetic (AFM) order is not essential for the decay, and that energy can be transmitted to general paramagnon excitations as long as short-range AFM spin correlations are present. FGR result can be quite well verified via a direct time 
evolution of the HD exciton decay when the perturbation term is switched on. Quite generally 
$\Gamma$ is well described with an exponential dependence 
\begin{equation}
\Gamma \sim \Gamma_0 \exp(-\alpha \Delta/J), \label{Gamma}
\end{equation}
obtained also by $n$th order perturbation theory arguments \cite{strohmaier10,sensarma10,sensarma11}  when considering the decay of unbound charged particles.
Since $\alpha$ involves parameters of the
model it is crucial for a fast recombination that within a MH insulator we find $\alpha <1$, 
being a consequence of the strong
charge-spin coupling. While one cannot treat the effective HD model analytically, we show that there is a very
helpful analogy with an exactly solvable exciton-boson (toy) model which confirms the form Eq.~(\ref{Gamma}), and moreover
allows direct interpretation of parameters, in particular $\alpha$. The final goal of this study is the comparison with experimentally
measured recombination rates in undoped cuprates NCO and LCO, and despite the fact that we propose only the minimal model for such process, obtained results 
are fairly close to the experimentally established ones\cite{okamoto10,okamoto11}.

The paper is organized as follows. In Sec.~II we present the derivation of the effective model from a single-band
Hubbard model via the canonical transformation. An analogous procedure is applied in Sec.~III to the three-band 
charge-transfer  model as directly relevant for undoped cuprates. Based on the existence of the 
bound HD exciton within the 2D effective model on a square lattice as established earlier, \cite{lenarcic13} 
we concentrate in Sec.~IV on the  calculation of recombination rate $\Gamma$ within the FGR approximation 
and on the comparison obtained with the direct time evolution.  In Sec.~V we present a 
toy exciton-boson model
within which decay rate $\Gamma$ can be evaluated 
exactly and even expressed analytically in the form analogous to Eq.~(\ref{Gamma}). 

\section{Single-band Hubbard model}\label{SecSingleB}

We start with the prototype model for the studies of the MH insulator - the single-band Hubbard model, 
\begin{equation}
H= -  t \sum_{ \langle ij \rangle s} (c^\dagger_{j s}  c_{i s} + {\rm H.c.})   
 + U \sum_i n_{i \uparrow} n_{i \downarrow}, \label{hub}
\end{equation}
where sum runs over nearest-neighbor (NN) pairs of sites $\langle ij\rangle$.
For the undoped cuprates the relevant lattice is 2D square lattice, which we
will consider further on.

We are interested in the half-filled case, $\bar n=1$, with a low density of holons
$\bar n_h \ll 1$ and doublons $\bar n_d \ll 1$. 
When discussing the recombination we would like to work with operators causing real, not just virtual transitions. 
To extract them we perform the usual canonical transformation of Hubbard model \cite{chao77,chao78,macdonald88} that in the lowest order decouples sectors with different number of HD pairs, however still relates them perturbatively. As shown later on, the transformed Hamiltonian in addition to the standard $t$-$J$ model \cite{dagotto94} contains also the terms causing recombination that were usually neglected in the studies of doped systems. Such effective model on one hand serves us to find the initial HD bound state by neglecting the recombination, and then yields its decay by taking it into account. One could perform also the transformation that completely decouples the sectors with different number of HD pairs\cite{macdonald88}, howevery this would not suit our purposes.

Hence, we rederive here the 
effective model employing  Hubbard operators $X_{i}^{pq}$, elaborated in Ref.\cite{ovchinnikov04}. If we define the 
holon state as $|H\rangle=|0\rangle$ and the doublon state as $|D\rangle =
c_{i\uparrow}^{\dagger} c_{i\downarrow}^{\dagger}|0\rangle$ operators are expressed as
\begin{align}
& X_{i}^{s H}=c_{is}^{\dagger}(1-n_{i\bar{s}}), 
\ X_{i}^{Ds}=-s c_{i\bar{s}}^{\dagger}n_{i s}, \ X_{i}^{DH}=s c_{is}^{\dagger} c_{i\bar{s}}^\dagger,
\notag \\
& X_{i}^{s\bar{s}}=c_{is}^\dagger c_{i\bar{s}},
\ X_{i}^{ss}=n_{is}(1-n_{i\bar{s}}), \label{EqX} \\
& \ X_{i}^{HH}=(1-n_{i\downarrow})(1-n_{i\uparrow}), 
\ X_{i}^{DD}=n_{i\downarrow}n_{i\uparrow}, \notag
\end{align}
where $s=\pm 1$ stands for the up/down electron spins. Upper incides $pq$ in $X_{i}^{pq}$ encode the initial (q) and final (p) state after the application of the operator.
In terms of the Hubbard operators the starting Hubbard model Eq.~(\ref{hub}) can be re-expressed as
\begin{align}
H&=H_{U}+H_{t}+H_{trc} =\notag \\
&=U\sum_{i}X_{i}^{DD} 
-t\sum_{ij,s}\left(X_{i}^{s H}X_{j}^{Hs}+X_{i}^{D\bar s}X_{j}^{\bar s D}\right) \notag \\
& \quad -t\sum_{ij,s}s\left(X_{i}^{s H}X_{j}^{\bar s D}+X_{i}^{D\bar s}X_{j}^{Hs}\right),
\end{align}
where $i,j$ are NN, and $H_U,H_t,H_{trc}$ are the on-site-repulsion, the HD-hopping and the HD-recombination/creation 
terms, respectively. 

\subsection{Canonical Transformation}
The canonical transformation is performed in the standard way \cite{chao77,chao78}
\begin{align}
\tilde H&=e^{S}He^{-S}=H+[S,H]+\frac{1}{2}[S,[S,H]]+\dots 
\end{align}
so that $H_{trc}$ is transformed out, consequently fixing $S$ with the condition 
$H_{trc}+ [S,H_{U}]=0$ to
\begin{equation}
S=\frac{t}{U} \sum_{ij,s}s\left(X_{i}^{sH}X_{j}^{\bar s D}-X_{i}^{D\bar s}X_{j}^{Hs}\right),
\end{equation}
and the transformed Hamiltonian up to second order in $t$
\begin{equation}\label{Ham 2}
\tilde H=H_{U}+H_{t}+[S,H_{t}]+\frac{1}{2}[S,H_{trc}].
\end{equation}
Using the $X$-operator commutation relations \cite{ovchinnikov04} we obtain several terms,
\begin{equation}
\tilde H= H_{tJ} + H_{rc} + H_{c}
\end{equation}
where  $H_{tJ}$ conserves the HD number
\begin{align}
H_{tJ}&=
-t \sum_{ij,s}X_{i}^{sH}X_{j}^{Hs}-t\sum_{ijs}X_{i}^{D\bar{s}}X_{j}^{\bar{s}D}+U\sum_{i}X_{i}^{DD} \notag \\
&+\frac{t^2}{U}\sum_{ij,s}(X_{i}^{s\bar s}X_{j}^{\bar s s} -X_{i}^{ss}X_{j}^{\bar s \bar s}), \label{EqHt}
\end{align}
and $H_{rc}$ is the essential term describing the HD recombination/creation
\begin{align}
H_{rc}= \frac{t^2}{U}\sum_{(ijk),s}s[&
X_{k}^{sH}(X_{i}^{ss}-X_{i}^{\bar s \bar s})X_{j}^{\bar sD}
+2X_{k}^{\bar sH}X_{i}^{s\bar s}X_{j}^{\bar sD} \notag \\
&+ {\rm H.c.} ], \label{EqHr1}
\end{align}
where $j,k$ are the NN sites to site $i$, and $j\neq k$. Further terms $H_c=H_4+H_5+H_6$ 
within the order $t^2/U$ are
\begin{align}
H_4 &=\frac{t^2}{U}\sum_{(ijk),s}s[
(X_{j}^{sH}X_{k}^{\bar s H}-X_{j}^{D\bar s}X_{k}^{Ds})X_{i}^{HD} \notag \\
&\qquad \qquad+X_{k}^{sH}X_{j}^{\bar{s}D}(X_{i}^{HH}-X_{i}^{DD})
+ {\rm H.c.}
] , \notag \\
H_5 &= \frac{t^2}{U}\sum_{(ijk),s}(-X_{j}^{sH}X_{i}^{DD}X_{k}^{Hs}
+X_{j}^{D\bar s}X_{i}^{HH}X_{k}^{\bar s D}  \notag \\
&- X_{j}^{s H}X_{i}^{HD}X_{k}^{Ds} +X_{j}^{Hs}X_{i}^{DH}X_{k}^{sD} 
-X_{j}^{sH}X_{k}^{Hs}X_{i}^{\bar s \bar s} + \notag \\
&+ X_{j}^{sH}X_{k}^{H\bar s}X_{i}^{\bar s s} +X_{j}^{D\bar s}X_{k}^{\bar s D}X_{i}^{ss} 
- X_{j}^{Ds}X_{k}^{\bar s D}X_{i}^{s\bar s}), \notag \\
H_6 & =\frac{t^2}{U}\sum_{ij}2(X_{i}^{DD}X_{j}^{HH}+X_{i}^{DH}X_{j}^{HD}) \label{EqHrMore}.
\end{align} 

Within the order $t^2/U$ the coupling between sectors with different number of HD-pairs is present 
in the terms $H_{rc}$, Eq.~(\ref{EqHr1}), and $H_4$, Eq.~(\ref{EqHrMore}). We note that
$H_4$ term could be relevant for recombination only at higher HD densities, since it is active 
only when three charged particles are NN to each other, being negligible at $\bar n_d, \bar n_h \ll 1$. 
Therefore it should not play a key role in the recombination at low density of holons and doublons discussed here,
and will be neglected further on. However, this term could be necessary for the description of short-time behavior in 
experiments where strong excitations produce an abundance of initially unbounded HD pairs. 
The terms  $H_5$ and $H_6$ only correct the excitonic wave functions  within the order $t^2/U$
and will also be neglected in comparison to the leading $H_{tJ}$, Eq.~(\ref{EqHt}).

\subsection{Effective Model}

The effective Hamiltonian that we consider further on contains terms Eqs.~(\ref{EqHt},\ref{EqHr1}). 
With the introduction of holon and doublon  creation and annihilation operators 
\begin{align}
h_{is}&=c_{is}^\dagger(1-n_{i\bar s})=X_i^{sH}, \notag\\
d_{is}&=c_{i\bar s} n_{is}=-s \ X_i^{sD},
\end{align} 
it can be written in a more compact and transparently spin-invariant way
\begin{align}
& H=H_{tJ}+H_{rc}\notag\\
& H_{tJ}=t\sum_{\langle ij\rangle,s}(h_{is}^\dagger h_{js} 
 - d_{is}^\dagger d_{js}  + {\rm H.c.})
 + U\sum_{i}n_{di} \notag \\
&\qquad +J \sum_{\langle ij \rangle}
\left(\mathbf{S}_{i}\cdot\mathbf{S}_{j}-\frac{1}{4}\delta_{1,n_{i}n_{j}}\right) \label{EqHtj}\\ 
& H_{rc}=t_{rc}\sum_{(ijk),ss'}\left(h_{ks}d_{js'}\vec{\sigma}_{s\bar{s}'}\cdot\mathbf{S}_{i} + {\rm H.c.} \right),\label{EqHrec}
\end{align}
where $n_{d}=(1/2)\sum_{is}d_{is}^\dagger d_{is}$ 
and $\vec\sigma=\{\sigma^x,\sigma^y,\sigma^z\}$ is a vector of Pauli matrices. Again $(ijk)$ signifies that $j,k$ are the NN sites to site $i$, and $j\neq k$. 
From the derivation we obtain that the recombination term, Eq.~(\ref{EqHrec}), has 
the coupling parameter $t_{rc}=2t^2/U=J/2$.

\section{Charge-transfer Hubbard model}

It is well known that on a microscopic level undoped and doped cuprates cannot be fully described within
the single-band Hubbard model, since they are undoped or doped Mott insulators of the charge-transfer type,
where more orbitals have to be included in the starting microscopic model.  
Therefore it is sensible to verify whether the recombination couplings obtained from the canonical transformation of 
the single-band Hubbard model are qualitatively correct approximation for the description of 2D cuprates.
We take the accepted multi-band tight-binding model for electrons on the 2D CuO$_2$ layers, 
including $3 d_{x^2-y^2}$ orbitals on Cu atoms and $2p_x/2p_y$ on O atoms. 
\cite{emery87,zhang88,zaanen88,ramsak89,muller-hartmann02}
In contrast to numerous  theoretical studies and models of hole doped systems, both type of charge carriers, 
positive an negative, have to be treated on the same level of approximation 
\cite{feiner96,raimondi96,tohyama04} in the present case
of excited MH insulator with holons and doublons.

\subsection{Multi-band Model}

 In the following, states are as usual (but in contrast to the previous section) defined relative to the filled 3d orbitals on copper and 2p orbitals on oxygen
 \cite{zaanen88}.  Including the NN Cu-O and O-O hopping, and the Coulomb repulsion on/between 
 Cu and O orbitals, the three-band $p$-$d$ model is written as  
\begin{align}
H=\sum_{i s} \epsilon_i n_{i s} + \sum_{\ave{ij} s} t_{ij}(c_{is }^\dagger c_{j s} + {\rm H.c.}) \notag \\
+ \sum_{i}U_i n_{i\uparrow} n_{i\downarrow} + \sum_{\langle ij\rangle} V_{ij} n_{i} n_{j}. \label{EqHbare}
\end{align}
Here $c_i$ (with corresponding $n_i$) stands for the annihilation of holes on different orbitals,
therefore equals either $c_i\equiv \bar d_i$ for $d$ orbitals with energy $\epsilon_d$ on copper at site $i$ 
or $c_i\equiv p_{xi}(p_{yi})$ 
for $p$ orbitals with energy $\epsilon_p$ on oxygen with positive displacement $x(y)$ relative to the NN copper at site $i$. 
We use notation $\bar d$ to avoid further confusion with doublon operators.
Hopping parameters equal $|t_{ij}|=t_{pd}, t_{pp}$ for hopping between NN
Cu-O and O-O orbitals, respectively, with sign dependent on the phases of facing orbitals.
Parameters $U_{i}=U_d, U_p$ take into account the on-site Coulomb repulsion on Cu and O orbitals, respectively,
while $V_{ij}=V_{pd}$ accounts for the repulsion between neighboring Cu-O orbitals. 
Introduced parameters have been extensively discussed in the literature. For numerical estimates further on
we use the concrete values $\epsilon_p-\epsilon_d=2.7, t_{pd}=1, t_{pp}=0.5, 
U_d=7, U_p=3, V_{pd}=1$, all in units of $t_{pd}\approx 1.3eV$, as used by others \cite{feiner96,muller-hartmann02}.

In the analysis we retain only a symmetrized oxygen orbital $(1/2)(|p_x\rangle -|p_y\rangle -|p_{-x}\rangle
+|p_{-y}\rangle)$, the one that hybridizes with the 
$d_{x^2-y^2}$ orbitals, leading to a two-band model \cite{zhang88,zaanen88,ramsak89}. 
Furthermore, we introduce their combinations - the orthonormal Wannier orbitals \cite{zhang88},
in framework of which the Hamiltonian can be separated into two parts: the local Hamiltonian $H_0$ describing the noninteracting cells, and 
the inter-cell coupling term $H_{cc}$. Each cell contains a Cu orbital and a Wannier O orbital.
Local part of the Hamiltonian has the form of a sum $H_0=\sum_i {\cal H}_{0i}$ of local intra-cell terms
\begin{align}
{\cal H}_{0i}&=\Delta_0 \sum_{s}n_{is}^p - \bar{t}_{pd} \sum_{\sigma}(\bar{d}_{is}^\dagger p_{is}+ {\rm H.c.}) \notag \\
&+ U_d \ n_{i\uparrow}^{\bar{d}} n_{i\downarrow}^{\bar{d}} 
+ \bar U_p \ n_{i\uparrow}^{p} n_{i\downarrow}^{p} + \bar V_{pd} \sum_{s s'} n_{is}^{\bar{d}} n_{is'}^p, \label{EqHOneCell}
\end{align} 
where $p_i^\dagger$ creates hole in the O Wannier orbital.
 Within the Wannier-orbital transformation parameters equal
$\Delta_0 =\epsilon_p-\epsilon_d-1.45~t_{pp}$, $\bar{t}_{pd} =1.92~t_{pd}, 
\bar U_{p}=0.21~U_p,  \bar V_{pd}=0.92~V_{pd}$, as taken from Ref. \cite{feiner96}.
In the inter-cell part $H_{cc}$ we retain only the dominant Cu-O and the O-O hopping,
\begin{equation}
H_{cc} =2 t_{pd} \mu_{10} \sum_{ijs} (\bar{d}_{i s}^\dagger p_{j s} + p^\dagger_{i s} \bar{d}_{js}) 
+2 t_{pp} \nu_{10} \sum_{ijs} p_{is}^\dagger p_{js}, \label{EqHcc}
\end{equation}
with coefficients $\mu_{10}=0.14, \nu_{10}=0.27$ for NN i and j sites, as taken from \cite{feiner96,raimondi96}.

\subsection{Local Charge and Spin States}

To discuss the recombination between holons and doublons we first have to identify states that 
represent them. Using the hole picture, doublon is represented by the filled Cu orbital, hence 
being the vacuum state $|D\rangle=|0\rangle$. On the other hand, holon is the 
generalized Zhang-Rice singlet \cite{zhang88,feiner96} $|H\rangle=H^\dagger |0\rangle$, obtained as the g.s. of local Hamiltonian 
${\cal H}_0$ in the singlet spin sector spanned by the states 
\begin{align}
\frac{1}{\sqrt{2}}(\bar d_{\uparrow}^\dagger p_{\downarrow}^\dagger - \bar d_{\downarrow}^\dagger 
p_{\uparrow}^\dagger)|0\rangle, \
\bar d_{\uparrow}^\dagger \bar d_{\downarrow}^\dagger|0\rangle, \
p_{\uparrow}^\dagger p_{\downarrow}^\dagger|0\rangle, \label{EqZR}
\end{align}
and has energy $E_{H}$. The single-hole state $|g_s \rangle$ (having correspondence to the spin background states in the single-band model) is calculated as the g.s. of ${\cal H}_0$ 
within the doublet sector spanned by 
\begin{equation}
\bar {d}_s^\dagger|0\rangle, \ p_{s}^\dagger|0\rangle \label{EqDoub}
\end{equation}
and has energy $E_g$. Besides the latter, we consider also the triplet states 
\begin{align}
&|T_{0}\rangle=\frac{1}{\sqrt{2}}(\bar d_{\uparrow}^\dagger p_{\downarrow}^\dagger 
+ \bar d_{\downarrow}^\dagger p_{\uparrow}^\dagger)|0\rangle, \notag \\
&|T_{-1}\rangle=\bar  d_{\downarrow}^\dagger p_{\downarrow}^\dagger|0\rangle, \
|T_{1}\rangle=\bar d_{\uparrow}^\dagger p_{\uparrow}^\dagger|0\rangle \label{EqStTrip}
\end{align}
with energy $E_T$. 
Other states, i.e. excited states within each sector, which can also be obtained with the diagonalization of $H_0$, will be neglected in our further analysis. Having higher energies they might be needed for the proper description of the early dynamics after the pump excitation, when highly excited states might be created. However, after the initial relaxation, we assume that system can be represented by the lowest lying states (which still represent also the excitations across the charge-transfer gap).

Although states $|H\rangle,|D\rangle,|g_{s}\rangle$ are a combinations of Wannier orbitals each of them is attributed to a single cell. Moreover, the hybridization between copper and oxygen orbitals, intrinsically present in them (as a consequence of basis vectors or diagonalization procedure), turns out essential when addressing the inter-cell hopping matrix elements of $H_{cc}$, Eq.~(\ref{EqHcc}), as discussed in App.~\ref{App3BandParams}. Still, they obviously bridge the single- and multi-band consideration by having analogues in the single-band picture.

\subsection{Reduced Hamiltonian}

We can now proceed by writing  the effective Hamiltonian in analogy with the single-band one by using the relevant states introduced in the previous subsection.
It is convenient to write the Hamiltonian with $X$ operators, analogously to those in the single-band model, 
Eq.~(\ref{EqX}), 
\begin{align}
&\bar X_i^{s D}= g_{is}^\dagger (1-n_{i}^{\bar d})(1-n_{i}^p), \\
&\bar X_{i}^{s H}=g_{is}^\dagger H_i, \
\bar X_{i}^{s T_{s'}}=g_{is}^\dagger T_{is'}, \notag \\
&\bar X_{i}^{s s}=g_{is}^\dagger g_{is}, \ 
\bar X_{i}^{s \bar{s}}=g_{is}^\dagger g_{i\bar{s}}, \notag \\
& g_{s}^\dagger=\cos\theta \ (1-n_{\bar{s}}^d)(1-n^p) \bar{d}_{s}^\dagger 
+ \sin\theta \ (1-n_{\bar{s}}^p)(1-n^d) p_{s}^\dagger  \notag
\end{align}
where $g_{is}^\dagger,H_i^\dagger,T_{is}^{\dagger}$ create the doublet g.s., holon (generalized Zhang-Rice singlet) and the triplet state, respectively. 
It still holds that $\bar X_{i}^{AB}=(\bar X_i^{BA})^\dagger$. Again $s=\pm1$ associated with $g_{is}$ stands for hole spin, whereas in $T_{is}$ it can have values $s=\pm1,0$ according to definitions in Eqs.~(\ref{EqStTrip}). To insure $X_i^{As}$ is nonzero only when applied to doublet g.s., its creation operator is written out explicitly, using parametrization elaborated in App. \ref{AppIntraStates}. In terms of such X operators we can present the Hamiltonian as the sum
$H=H_t+H_{trc} + H_{dg}$, representing the   effective HD hopping  
(containing possible creation of triplet states), their recombination and the diagonal part, respectively,
\begin{align} 
H_t&=  \sum_{ij ,s=\pm 1} \big( t^h \bar X_i^{Hs} \bar X_j^{sH} 
+ t^d \bar{X}_i^{sD} \bar{X}_j^{Ds}\big)  \label{EqHt3}\\
&+\sum_{ij ,s=\pm 1}\big( - s \ t^{T_0} \bar{X}_i^{s H} \bar{X}_j^{T_{0}s}
+s \ t^{T_1} \bar{X}_i^{\bar{s} H} \bar{X}_j^{T_{s}s} + {\rm H.c.} \big) \notag  \\
H_{trc}&=\sum_{ij, s=\pm 1}
\big(-s \ t^r \bar{X}_i^{\bar{s}H} \bar{X}_j^{s D}
+ t^{r_0} \bar{X}_i^{\bar{s} T_0} \bar{X}_j^{s D} \label{EqHrc3}  \\
& \quad\quad\quad\quad +t^{r_1} \bar{X}_i^{ sT_{s}} \bar{X}_j^{s D} + {\rm H.c.}  \big) \notag \\
H_{dg}&=\sum_{i} \big(  \epsilon_H \bar{X}_i^{HH} + \epsilon_D \bar{X}_i^{DD} 
+\epsilon_T \sum_{s=\pm1,0} \bar{X}_i^{T_sT_s}\big), \label{EqHdg3}
\end{align}
where $i,j$ are NN. Values $\epsilon_H=E_H-E_g,\epsilon_D=-E_g,\epsilon_T=E_T-E_g$ are the single-cell energies of holon, doublon and 
triplet relative to the doublet g.s., respectively. Dependence of the introduced couplings 
$t^c, c=h,d,T_0,T_1,r,r_0,r_1$ and energies $\epsilon_{H},\epsilon_{D}, \epsilon_{T}$ on the parameters of the original Hamiltonian 
Eqs.~(\ref{EqHOneCell},\ref{EqHcc}) is presented in the App. \ref{App3BandParams}.

\subsection{Effective Hamiltonian}

Similarly to the treatment of the single-band Hubbard model within the $U\gg t$ limit in Sec.~II
we transform out the recombination/creation term  $H_{trc}$ with a canonical transformation 
$e^S H e^{-S}$. Operator $S$ is determined by the condition $[S,H_{dg}]+H_{trc}=0$. 
After the transformation, HD recombination/creation term $H_{rc}$ again acts between the next-NN cells, however,
now one has to distinguish between channels leading to different configurations of spins in the doublets of final state, since their amplitudes $r^i$ are different
\begin{align}
& H_{rc}= -\sum_{(ijk),s}s[
\bar{X}_{k}^{sH}(r^h \bar{X}_{i}^{ss}- r^d \bar{X}_{i}^{\bar s \bar s})\bar{X}_{j}^{\bar sD} \notag \\
&\qquad\qquad\qquad+r^{hd} \bar{X}_{k}^{\bar sH}\bar{X}_{i}^{s\bar s}\bar{X}_{j}^{\bar sD}
+ {\rm H.c.} ],\label{EqHrc3fin}\\
& r^{h}=\left(\frac{ t^h t^r}{\epsilon_H+\epsilon_D} + \frac{t^{T_0} t^{r0}}{\epsilon_T+\epsilon_D}\right), \notag\\
& r^d=\left(\frac{t^d t^r}{\epsilon_H+\epsilon_D} - 
\frac{t^{T_1} t^{r1}}{\epsilon_T+\epsilon_D}\right), \label{EqRd} \\
& r^{hd}=\left(\frac{(t^d + t^h) t^r}{\epsilon_H+\epsilon_D} - \frac{t^{T_0} t^{r0}}
{\epsilon_T+\epsilon_D}\right).\notag 
\end{align}
Not only different amplitudes of holon and doublon hopping parameters, but also new processes of recombination via intermediate triplet states alter the result. To obtain the latter, hopping terms involving triplet states were included in $H_t$ and $H_{trc}$ in the first place. Although they exhibit reacher physics of multi-band model, one should be aware that recombination via triplet state causes only smaller corrections in the coupling strengths, since $\epsilon_H \ll \epsilon_T$.
However, pure form of Eq.~(\ref{EqHrc3fin}) is very similar to its single-band analogue Eq.~(\ref{EqHr1}) with an additional overall minus that is a consequence of transition from electron to hole picture.

\begin{figure}[ht]
\includegraphics[width=0.43\textwidth]{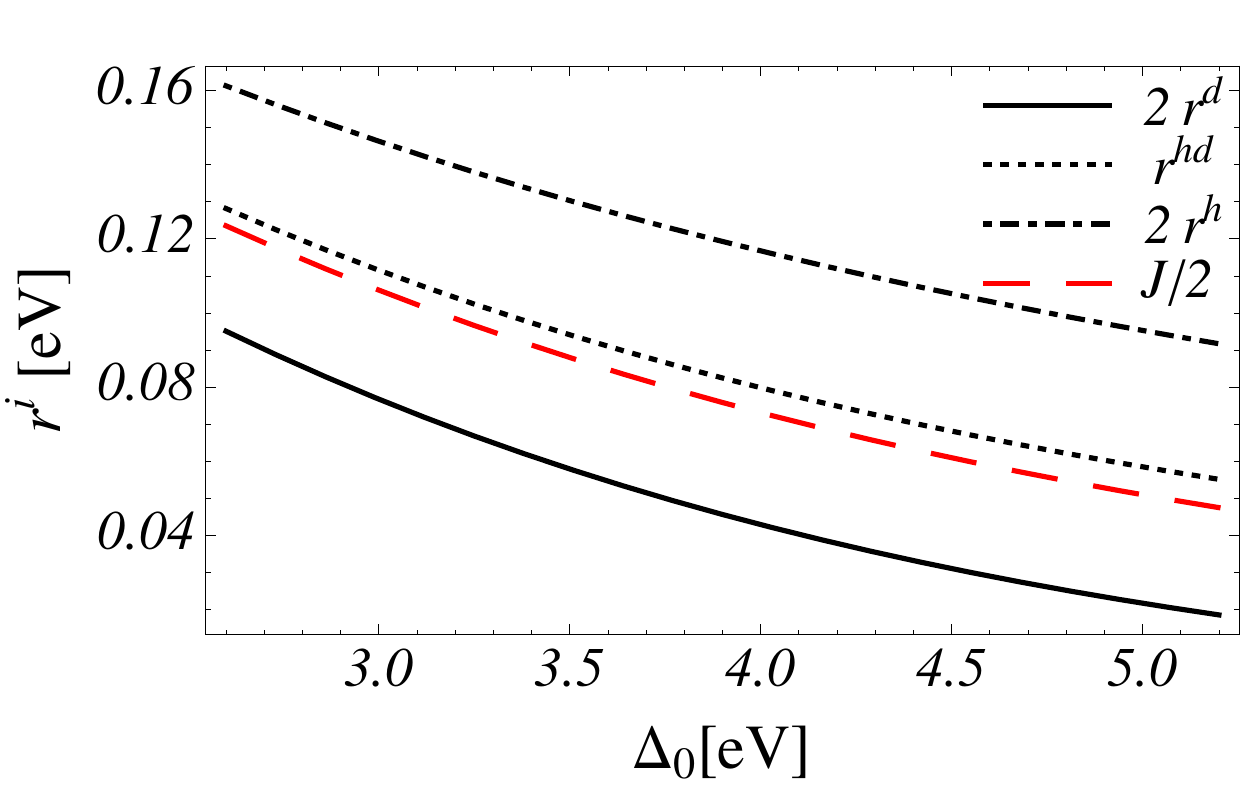}
\caption{(Color online) Comparison of coupling parameters $r^{i}=2r^{d},r^{hd},2r^{h}$ for different recombination channels with the (rescaled) exchange coupling $J/2$ as a function of charge-transfer gap $\Delta_0$. For other parameters standard values are used. }
\label{FigTrec}
\end{figure}

If we calculate all three relevant recombination couplings $2r^d, r^{hd},2r^h$ at realistic parameters we confirm that they are not far away from $t_{rc}=J/2$, the value obtained from the single-band model. Their dependence on $\Delta_0$ is plotted in Fig. \ref{FigTrec}. Rescalations are made for clearer comparison with $J/2$. Using the same procedure via intra-cell diagonalization, exchange coupling plotted is expressed as \cite{feiner96}
\begin{equation}
J=4 \left(\frac{(t^r)^2}{\epsilon_H+\epsilon_D} -\frac{(t^{r_0})^2}{\epsilon_T+\epsilon_D}\right).
\end{equation}

To exhibit the spin invariance of $H_{rc}$ we define (as in the single-band model) 
$\tilde{d}_{is}=-s\bar{X}_i^{sD},\tilde{h}_{is}=\bar{X}_{i}^{sH}$ 
in term of which $H_{rc}$ obtains a form similar to Eq.~(\ref{EqHrec}),
\begin{align}
H_{rc}= -\sum_{(ijk) ss'} & \left[ \tilde{h}_{is} \tilde{d}_{ks'}  \left( 
r^{hd}\vec{\sigma}_{s\bar{s}'} \cdot {\bf S}_j + \bar r^{hd} \mathbb{1}_{s\bar{s}'} 
\right) + {\rm H.c.} \right],  \label{EqRecCT} 
\end{align}
where we used $\bar r^{hd} = (r^{h}-r^{d})/2$ and
 $t^{T_1} t^{r1}=2t^{T_0} t^{r0}$, see App.~\ref{App3BandParams}.


\section{Exciton recombination rate}\label{SExcRecRate}

In previous Secs.~II, III it was shown that both the single-band Hubbard model as well 
as the three-band model for cuprates reduce at low HD density to the same 
effective model with the only difference being the strengths of the recombination/creation terms in $H_{rc}$.

\subsection{Holon-doublon Exciton}

In order to explain the experimentally observed independence of decay rate $\Gamma$ 
on the pump fluence, i.e. also the exponential decay of HD density,
we first have to determine the existence of the bound HD pair.
This is based on argumentation that if pairs were not bound, recombination process would depend on the probability to encounter the oppositely charged particle, evidently leading to a non-exponential decay (unless thermal charge density is high). 
Present problem of HD binding has analogies with binding of holes in doped cuprates, also studied withing the $t$-$J$ model \cite{dagotto94,chernyshev98}. Although the origin of binding is in both cases the same, indistinguishable two holes $N_h=2$ form a $d$-type bound state, whereas the distinguishable doublon and holon form a 
$s$-type (A$_1$ symmetry) bound pair, which is indeed found numerically
\cite{tohyama06,lenarcic13}. Since latter state has even symmetry it is not accesible by optical transition from the insulator AFM state. On the other hand, the optically active $p$-type state with binding energy $\epsilon_b  \gtrsim 0$ within our calculation does not seem to be a bound one . 

Knowing that at low charge density coupling between sectors with different number of HD pairs is weak, we first neglect the recombination/creation term $H_{rc}$ that causes transitions between sectors, and extract the initial HD state $|\psi^{hd}_0 \rangle$ from the spectrum of eigenstates of $H_{tJ}$, Eq.~(\ref{EqHtj}), as the g.s. in the sector with one HD pair. 
Calculating it in the single HD pair sector for system of limited size we neglect possible interaction between different pairs, justified for the cases of low charge density.

\begin{figure}[ht]
\includegraphics[width=0.25\textwidth]{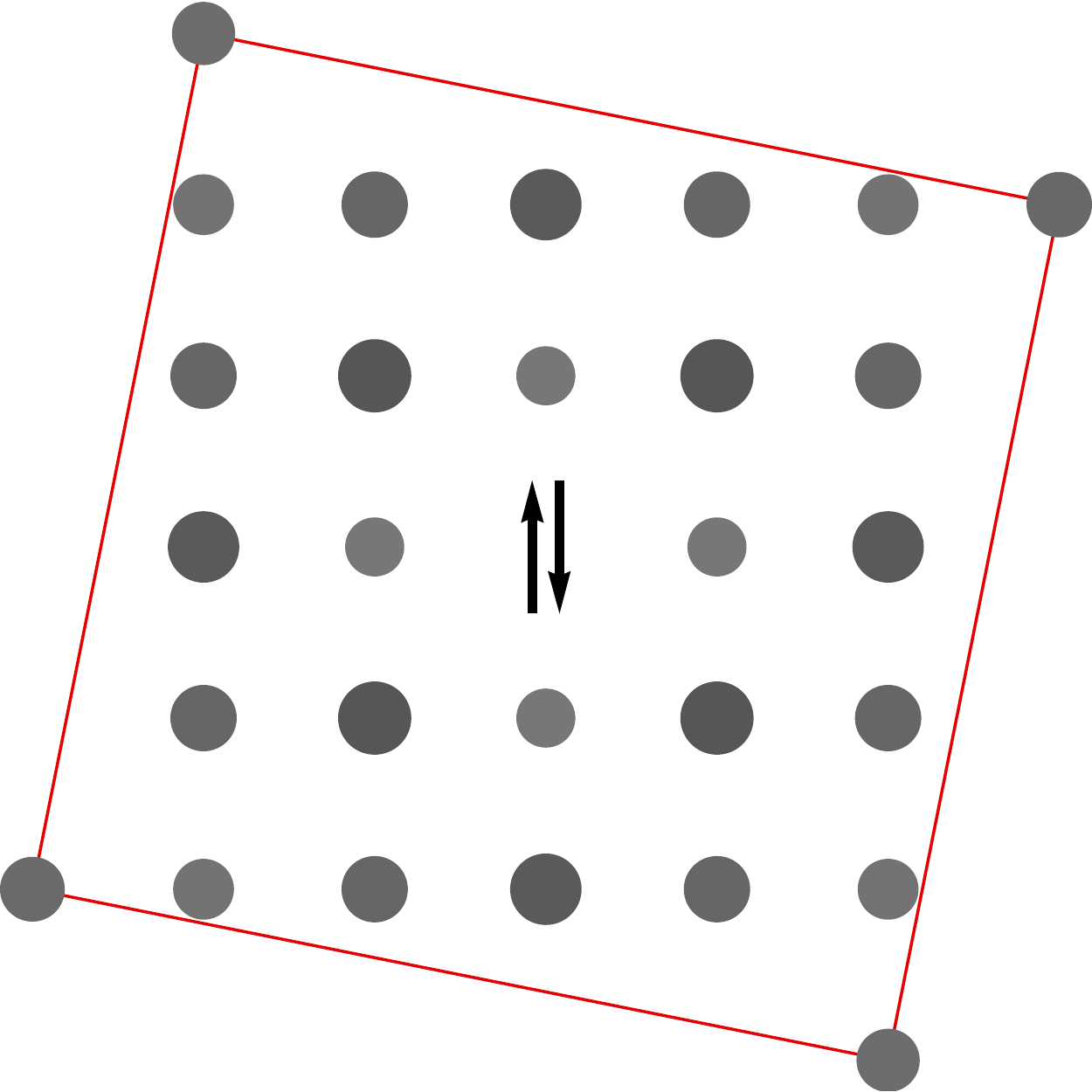}
\caption{(Color online) Charge density correlation $D_j$. }
\label{FigDistrib}
\end{figure}

Binding properties of HD state $| \psi^{hd}_0 \rangle$
were obtained via exact diagonalization of $H_{tJ}$ using the Lanczos technique on 
the square lattices with $N \leq 26$ sites and periodic boundary conditions. 
Here we shall skip the detailed analysis and results presented in Ref. \cite{lenarcic13}. In short, we calculated the HD binding energy $\epsilon_b= E^{hd}_0 -E^h_0 -E^d_0+ E^0_0$
where $E^{hd}_{0},E^h_0, E^d_0, E^0_0$ correspond to the HD pair, single hole, single doublon and the undoped AFM g.s., respectively. 
In the regime of interest for cuprates ($J/t=0.3-0.4$) the lowest ($s$-type) state shows appreciable binding $\epsilon_b/t \sim - 0.4$, quite robust towards the finite size effects \cite{lenarcic13}.
It should be pointed out that the inclusion of longer-range Coulomb repulsion would even enhance 
$|\epsilon_b|$  but is not expected to be the driving or dominant effect (results presented in 
Ref.~\cite{lenarcic13}) in the 2D square lattice. 
As an additional proof of HD binding we calculate also the exciton density correlations
$D_j=\langle \psi^{hd}_0| n_{hj} n_{d0}  | \psi^{hd}_0 \rangle$
(for the purpose of presentation the position of doublon is chosen as the origin). $D_j$ 
obtained on $N=26$ for $J=0.4$ are presented in Fig.~\ref{FigDistrib}, showing consistence with the binding since HD pair is with the largest probability on a distance $d_0 = \sqrt{2}$, as is also the case for the $d$-wave hole 
binding within the 2D $t$-$J$ model \cite{dagotto94,chernyshev98}.  

\subsection{Recombination Rate via Fermi Golden Rule}

The HD exciton $| \psi^{hd}_0 \rangle$ is not an eigenstate of the effective model when perturbation $H_{rc}$, Eq.~(\ref{EqHrec}), is included.
A standard approach to evaluate the decay rate into a continuum of states is the 
Fermi golden rule,
\begin{equation}
\Gamma=2\pi \sum_{m}|\langle\psi_{m}^0|H_{rc}|\psi_{0}^{hd}\rangle|^2 \ 
\delta(E_{m}^{0}-E_{0}^{hd}), \label{EqGam}
\end{equation}
where the matrix elements are highly nontrivial since they represent the overlap of 
modified exciton wave function  $H_{rc}|\psi_{0}^{hd}\rangle$ on 
highly spin-excited (multi-magnon) states $|\psi_{m}^0\rangle$ with energy $E_m^0$ within the undoped AFM spin system.
Our application of the FGR approximation has many analogies, recently employed in the analysis of the  
decay of excitons via multi-phonon emission in nanotubes \cite{avouris06,perebeinos08}. 
For the numerical consideration it is crucial that Eq.~(\ref{EqGam}) can be represented as a resolvent 
$\Gamma= -2~ {\rm Im} C(\omega=\Delta)$, where $\Delta =E_0^{hd}-E_0^0$
is the excitation gap, and
\begin{equation}
C(\omega)=\langle\psi_0^{hd}|H_{rc} \frac{1}{\omega^+ +E_0^0 -H_{J} }H_{rc}|
\psi_0^{hd}\rangle, \label{EqC}
\end{equation}
with $\omega^+=\omega+i \delta$. In the evaluation only the exchange part
$H_J$ of the  $H_{tJ}$, Eq.~(\ref{EqHtj}), is relevant. 

\begin{figure}[ht]
\includegraphics[width=0.43\textwidth]{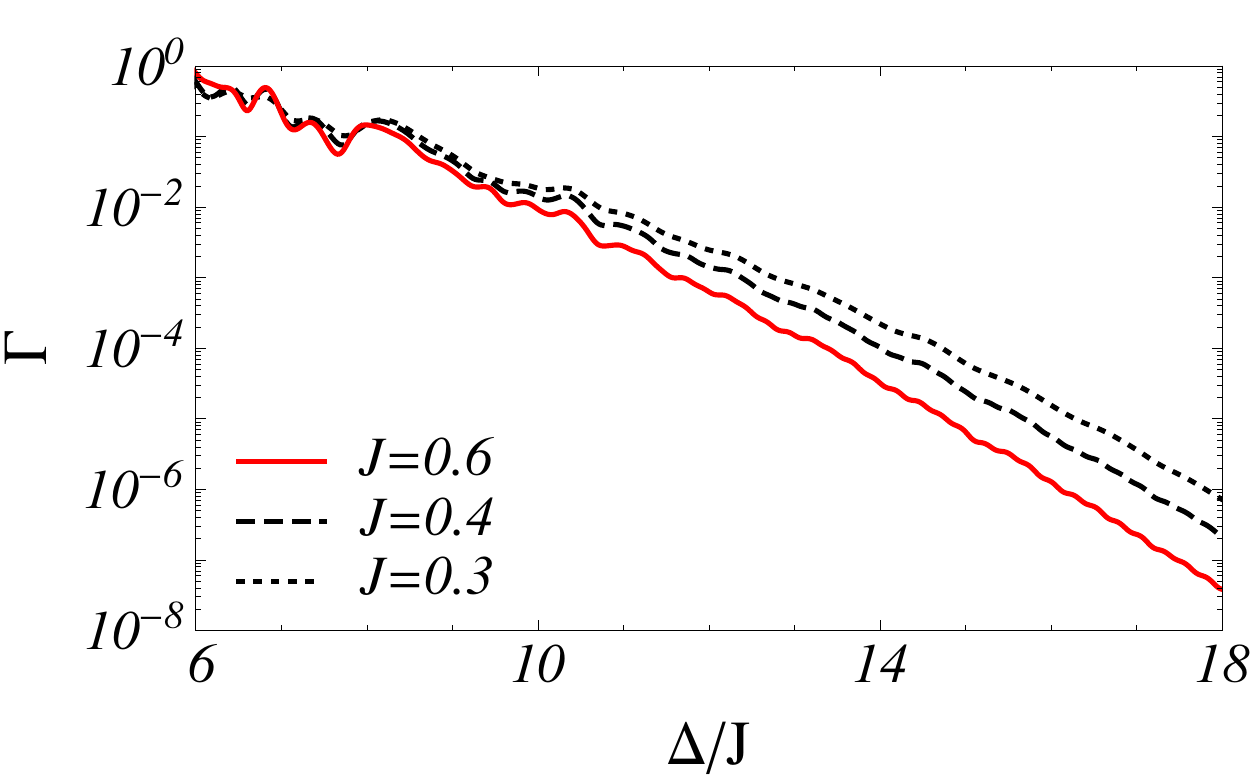}
\caption{(Color online) Exciton recombination rate $\Gamma$ vs. $\Delta/J$
for different $J=0.3,0.4,0.6$ as calculated for $N=26$ sites.}
\label{fig4}
\end{figure}

Within Lanczos procedure Eq.~(\ref{EqC}) can be evaluated\cite{dagotto94,prelovsek11} on 2D square lattice with up to $N=26$ sites\cite{lenarcic13}.
In Fig.~\ref{fig4} the dependence $\Gamma(\Delta)$ for $J=0.3,0.4,0.6$ is presented. Here the energy of HD pair $\Delta$ that has to be transmitted to the spin excitations, $\Delta=E_m^0-E_0^0$, is taken as a parameter independent of $J$. 
As suggested from Fig.~\ref{fig4} decay rate $\Gamma$ shows approximately exponential dependence on $\Delta/J$, Eq.~(\ref{Gamma}), with effective $\alpha$ in the range $0.3 <\alpha <0.7$ (for chosen $0.3\le J \le 0.6$). This signals that there is some additional subtle $J$ dependence, besides the exponential dependence on the number of spin excitations $n\sim \Delta/J$ created. 

As discussed already in Ref.~\cite{lenarcic13} essential ingredient for the substantial decay is dressing of HD pair with spin excitations, revealed by deviations in bond energy of the exciton state relative to the AFM g.s\cite{lenarcic13}. In the process of recombination this local spin perturbation can be even enhanced, and finally has to disperse into the whole system.
An attempt to relate both aspects is to motivate the dependence of decay rate on $\Delta$ and $J$ via the construction of sufficient spin dressing of cca. $n$ spin flips as a $n$-th order perturbation process \cite{lenarcic13}, as suggested by previous similar considerations \cite{strohmaier10,sensarma10,sensarma11}. According to these arguments, following from the appropriate matrix element, decay rate should have the form
\begin{equation}
\Gamma \propto {\rm exp}\left[ - \alpha_0 \frac{\Delta}{J} \ln{\frac{\Delta}{e t} }\right] . \label{fit2}
\end{equation} 
with $\alpha_0=2$. However, when fitting Eq.~(\ref{fit2}) to the numerical data, $\alpha_0\approx 0.8$ with modest $J$ dependence is obtained\cite{lenarcic13}. In Ref.\cite{sensarma11} the additional structure of constant $\alpha_0$ was treated with self-avoiding path reasoning, though not for the bound HD pair.
Our more elaborate, however not necessarily unrelated consideration of charge-spin coupling using exciton-boson model will be given in the next section.

\subsection{Recombination Rate via Direct Time Evolution}

\begin{figure}[ht]
\includegraphics[width=0.43\textwidth]{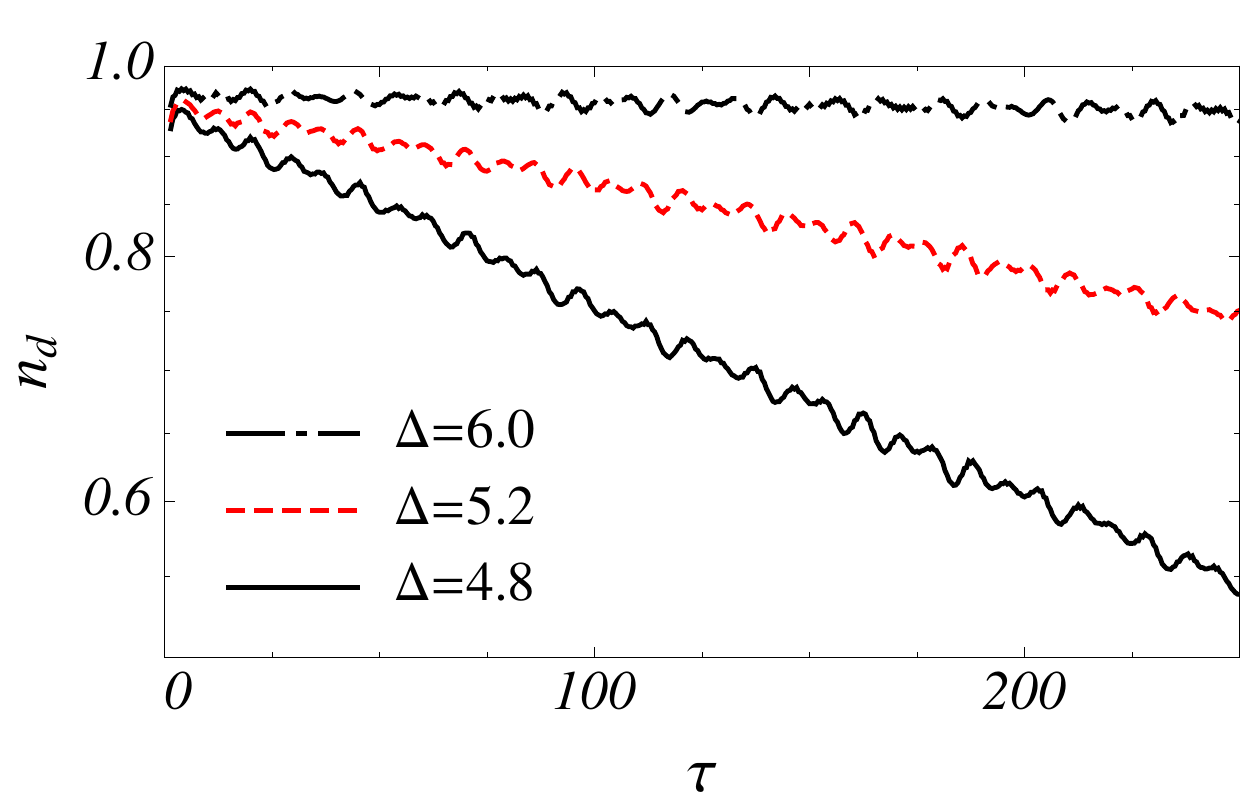}
\caption{(Color online) Doublon (and also HD pair) occupation number $n_{d}$ (in logarithmic scale) 
as a function of time $\tau$, 
calculated for different gaps $\Delta=4.8, 5.2, 6.0$ and parameters $J=0.4$ for system of size $N=26$.}
\label{figsup1}
\end{figure}

In order to validate the approximation using the FGR, Eq.~(\ref{EqGam}), 
we perform also direct time evolution of the same initial excitonic state $|\psi_{0}^{hd}\rangle$ 
under the whole Hamiltonian $H=H_{tJ}+H_{rc}$, however, we restrict the Hilbert space only to the sectors with zero and one HD pair. 
In Fig.~\ref{figsup1} we present the time evolution of the doublon (also the HD pair) occupation number, 
\begin{equation}
n_{d}(\tau)=\frac{1}{2}\langle \psi(\tau)|\sum_{is}d_{is}^{\dagger}d_{is}|\psi(\tau)\rangle.
\end{equation} 
The evolution of $|\psi(\tau)\rangle$ is obtained by solving the time-dependent 
Schr\"odinger equation using the
Lanczos method \cite{park86,prelovsek11}. In Fig.~\ref{figsup1} we present and compare results for $J=0.4$ and different effective gaps 
$\Delta=4.8,5.2,6.0$, as calculated for the system with $N=26$ sites. 
Effective gap is defined using $|\psi_{gs}\rangle$ (g.s. of $H$ within our restricted Hilbert space) as
\begin{equation}\label{EqDeltaTime}
\Delta=\ave{\psi_0^{hd}| H | \psi_0^{hd}} - \ave{\psi_{gs}|H| \psi_{gs}},
\end{equation}
since it turns out to be a function of the coupling strength $t_{rc}$ due to adiabatic change of the eigenspectra of $H$ caused by $H_{rc}$. By adiabatic we mean that even though the whole energy of each eigenstate is shifted, the fraction of spin excitations within it is preserved, and it is the amount of spin excitations that should label the final states when discussing the recombination.
Rapid oscillations seen in Fig. \ref{figsup1} emerge due to fast switching of $H_{rc}$ and finite-size effects, however, they get evidently reduced with bigger $N$. For clarity averaging over $\delta \tau=3$ is used.

\begin{figure}[ht]
\includegraphics[width=0.43\textwidth]{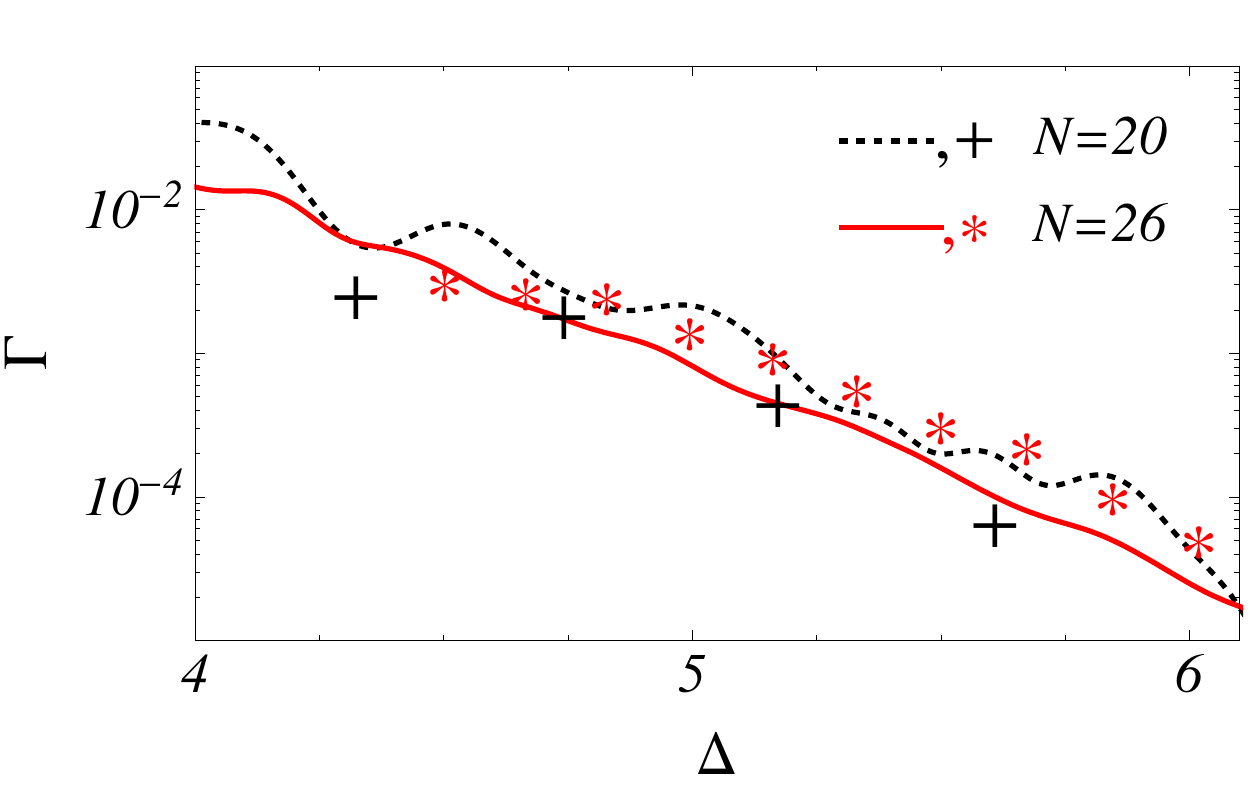}
\caption{(Color online)  Comparison of the exciton recombination rate $\Gamma$ vs gap $\Delta$ as calculated using the FGR (lines) and time evolution (dots) for $J=0.4$ and systems of size $N=20,26$.}
\label{figsup2}
\end{figure}

From  Fig.~\ref{figsup1} we can confirm that after an initial transient an exponential decay  is established. 
When simulating recombination on a finite system one should be aware that the finite-size level distance $\delta\omega$ limits
the long-time evolution to $\tau\approx 2\pi/\delta\omega$, and is for system with $N=26$ sites of order
$\delta\omega \approx 10^{-1}$.
Using the fit $\log n_{d}(\tau)=-\Gamma \tau + \log n_{d0}$, one can compare the result obtained for 
$\Gamma$ with the one calculated with FGR. Fig.~\ref{figsup2} shows this comparison for 
$J=0.4$ and system sizes  $N=20,26$. Lines correspond to the result from FGR, while dots are obtained from the fits to $\log n_{d} (\tau)$ in the span of interesting $\Delta$. 
We obtain a quite good agreement between the two methods, as shown in Fig.~\ref{figsup2}. 
Both methods confirm the exponential dependence Eq.~(\ref{Gamma}). Somewhat smaller $\Gamma$ obtained with time evolution on $N=20$ lattice could be attributed to the decay into the discrete 
multi-magnon spectra, which is sparser at smaller lattices. 


\section{Coupled exciton - boson model}

Our numerical results clearly reveal approximate exponential dependence of decay rate
$\Gamma$, Eqs.~(\ref{Gamma}), on the number of  bosonic excitations $n \sim \Delta/J$
created in the recombination process. As mentioned in the previous section such dependence has been reproduced qualitatively also 
via the $n$-th order perturbation arguments \cite{strohmaier10,sensarma10,sensarma11,
lenarcic13}, yet the constant $\alpha_0\approx 0.8$ obtained from fiting Eq.~(\ref{fit2}) to numerical results cannot be given a clear origin. It would be desirable to have a solvable model, which 
could qualitatively or even quantitatively simulate the observed physics.  
Relying on the interpretation developed in the previous section; suggesting that in the process of recombination spin excitations dressing the HD pair are dispersed into the whole system,
it seems plausible to formulate the problem more generally - as a decay of an excitonic 
state $|e\rangle=e^\dagger|0\rangle$ due to coupling to bosonic degrees of freedom. 
The main physics of such formulation can be captured with an exciton-boson toy model,  
used on a similar problem to interpret the charge recombination in carbon nanotubes via the multi-phonon emission \cite{avouris06,perebeinos08},
here generalized to dispersive bosons,
\begin{align}
H=& \ H_e+H_{eb}+H_{b}+H_{rc} \label{EqHtoy}  \\
= &\ E_{e}e^{\dagger}e + e^\dagger e\sum_{q}\lambda_q (a_q^\dagger + a_{-q}) + 
\sum_{q}\omega_q a_q^\dagger a_q + \notag \\
&+ g_{rc}(e+e^\dagger). \notag
\end{align}
$a_q^\dagger$ is creation operator for bosons with momentum $q$ and energy $\omega_q$. 
The exciton-boson coupling is mediated by the term  $H_{eb}$, while $H_{rc}$
represents the simplest form of the exciton recombination/creation. It is evident that such model only
indirectly simulates the full physics of exciton coupled to spin fluctuations, Eqs.~(\ref{EqHtj},\ref{EqHrec}).

The toy model Eq.~(\ref{EqHtoy}) basically describes the two-level system coupled to bosons, 
and was used when discussing related question of radiationless transitions in large molecules \cite{englman70}, quantum dissipation \cite{leggett87}, and in numerous other problems.
It is well analyzed and solvable in several limits, in particular if $H_{rc}$ is treated as a perturbation.

Drawing analogies with procedure in the previous section, we would like to obtain the excitonic wave function dressed with bosons and get rid of the strong coupling between exciton and bosons on the level of unperturbative part of the Hamiltonian. Therefore we do the standard
Lang-Firsov transformation $\tilde H=e^{-S}He^{S}$, which eliminates $H_{eb}$ with 
\begin{equation}
S=- e^\dagger e\sum_{q} \alpha_q (a_q^\dagger - a_{-q}),
\end{equation}
where $\alpha_q= \lambda_q/\omega_q$ and yields the transformed Hamiltonian
\begin{align} 
&\tilde H=\tilde{H}_0 + \tilde{H}_{rc}, \notag \\
&\tilde{H}_0=(E_{e}-\epsilon_{eb}) e^\dagger e 
+ \sum_q \omega_q a_q^\dagger a_q, \label{EqH0Trans}\\
&\tilde{H}_{rc}= g_{rc} \exp{[-\sum_q \alpha_q(a_{q}^\dagger - a_{-q})]}~e  + {\rm H.c.}. \notag
\end{align}
The exciton-boson binding energy $\epsilon_{eb} = \sum_q \lambda_q^2/\omega_q$ 
that lowers the exciton's energy implicitly indicates its bosonic dressing. However, it is assumed to be modest, i.e. $\epsilon_{eb}\ll E_e$. As before the initial wave function is obtained, neglecting $\tilde H_{rc}$, as the ground state of $\tilde{H}_0$ in the sector with an exciton 
$|\psi_0 \rangle=e^\dagger |0\rangle$, having energy $E_e-\epsilon_{eb}$. 

Switching on $\tilde{H}_{rc}$ the exciton starts to decay  and we evaluate the recombination rate $\Gamma$
using the FGR again, now written in form of an integral
\begin{align}
\Gamma
&=-2Im
\ \langle \psi_0 | 
\tilde H_{rc} \ \frac{1}{\omega + E_{\tilde 0}-\tilde H_0} \ \tilde H_{rc} 
|\psi_0\rangle \label{EqGamma0} \\
&=2Im \ i \ \langle \psi_0 | \tilde H_{rc} \int_0^\infty dt e^{i\omega t} 
\ e^{-i (\tilde{H}_0 - E_{\tilde 0})t} \ \tilde H_{rc} |\psi_0 \rangle. \notag
\end{align}
where $E_{\tilde{0}}$ is the g.s. energy in the sector without exciton. Taking into account well known relations for coherent states 
(since $\tilde{H}_{rc}|\psi_0\rangle$ is a coherent state)
 \begin{equation}
\langle \psi_0 | \tilde H_{rc} e^{-i(\tilde{H}_0 - E_{\tilde 0}) t} \ \tilde H_{rc} | \psi_0 \rangle
= g_{rc}^2 \exp{[\sum_q \alpha_q^2(e^{-i\omega_q t}-1)]},\label{EqCohMat}
 \end{equation}
we finally get
\begin{equation}
\Gamma =2g_{rc}^2  \ Re \int_0^\infty dt 
 \exp{[ i\omega t + \sum_q \alpha_q^2(e^{-i\omega_q t }-1) ]}. \label{EqGammaInt}
\end{equation}
Here $\Gamma$ should be evaluated at 
$\omega=E_{e}-\epsilon_{eb}$,
which is the difference in the g.s. energy of $\tilde H_0$ in the sector with and without the exciton. 
 
\subsection{Saddle point approximation}

While Eq.~(\ref{EqGammaInt}) can easily be evaluated numerically for 
arbitrary parameters, i.e. the coupling strength $g_{rc}$ and dispersions $\lambda_q, \omega_q$,
it is instructive to get result in 
a form that reveals the relevant quantities entering $\Gamma$. For this purpose
we first simplify the general dispersions $\lambda_q, \omega_q$ by
assuming that the boson coupling function $g(\omega)$ has mean energy $\omega_0$ and a 
$\sigma$ spread around that value, fixing the form
\begin{equation} \label{EqXi}
g(\omega) = \sum_q \alpha^2_q \ \delta(\omega-\omega_q)
=\frac{\xi}{\sqrt{2\pi} \sigma}  \ e^{-(\omega-\omega_0)^2/2 \sigma^2}
\end{equation}
with a Gaussian function centered at $\omega=\omega_0$. 
The dimensionless prefactor $\xi=\sum_q \alpha_q^2$ takes into account the strength of the coupling.
Such approximation is well justified for bosons with weak dispersion, e.g. the optical phonons,
however it should be reasonable also for the 2D magnons under examination with $\omega_0\approx J$. 
Nevertheless, the dispersion $\sigma>0$ is 
essential for smooth variation of $\Gamma$ vs $\omega$, and conceptually crucial for final dispersion of bosons into the system. 

\begin{figure}[ht]
\includegraphics[width=0.40\textwidth]{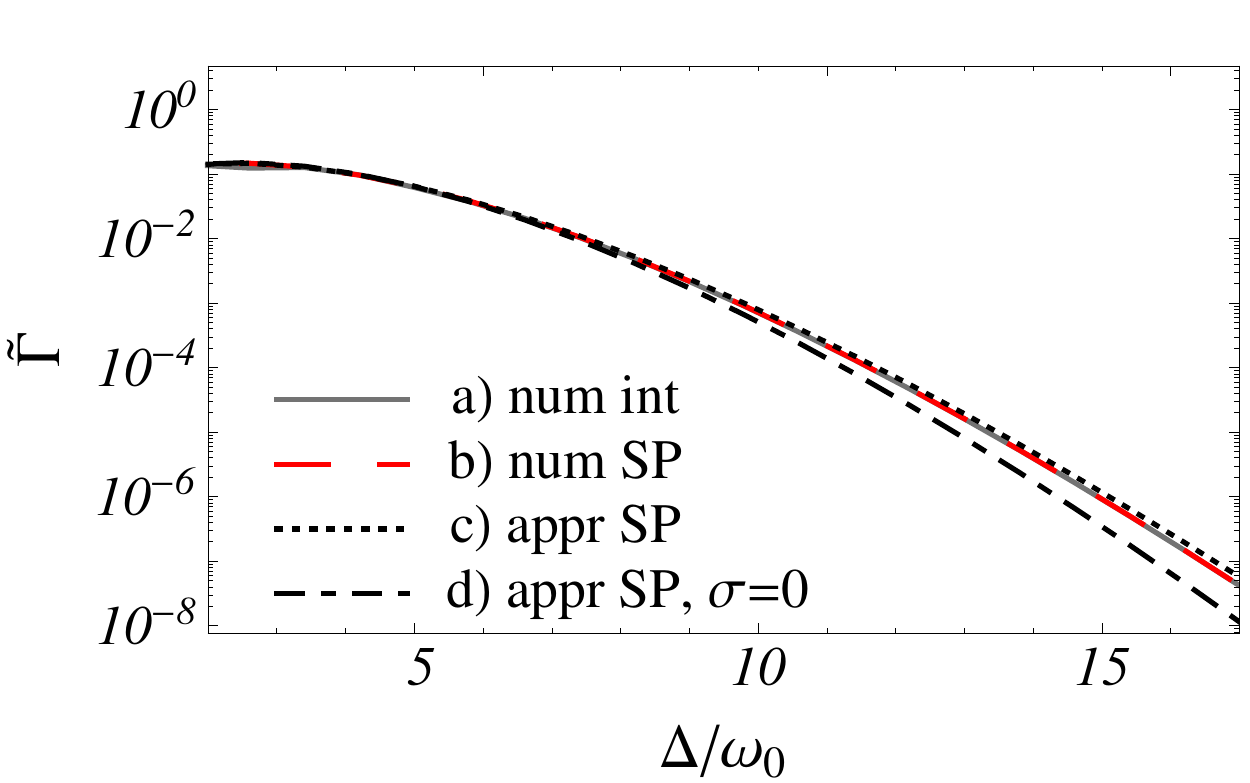}
\caption{(Color online) Comparison of the result for $\tilde\Gamma=\Gamma/g_{rc}^2$, if calculated with a) the numerical evaluation of $\Gamma$ from Eq.~(\ref{EqGammaInt}), b) the saddle point result for numerically (exactly) established saddle, c) the saddle point result for approximate saddle Eq.~(\ref{EqSPp}), and d) compact form of Eq.~(\ref{EqSPGama}). Parameters $\omega_0=5,\xi=3,\sigma=\omega_0/4$ are used so that numerical integration a) is well defined.}
\label{FigSPInt}
\end{figure}

The advantage of the form Eq.~(\ref{EqXi}) is that the integral Eq.~(\ref{EqGammaInt}) can be analytically 
evaluated by saddle point method \cite{morse1953book}, i.e.
\begin{align}
\int_{-\infty}^{\infty} e^{f(t)} dt
\approx e^{f(t_0)} \sqrt{\frac{2\pi}{-f''(t_0)}}, \quad f'(t)|_{t_0}=0. \label{EqSP0}
\end{align}
The function $f(t)$ and its saddle point $t_0$, correct up to $\mathcal{O}(\sigma^4/\omega_0^4)$, are in our case
\begin{align}\label{EqSPp}
& f(t)=i\omega t+ \xi \  e^{-i\omega_0 t - \sigma^2 t^2/2}, \notag \\
 & t_0=\frac{i}{\omega_0 +\tilde{\sigma}} 
 \ln\left( \frac{\omega}{\xi(\omega_0+2\tilde{\sigma})}\right),
\end{align}
where $\tilde\sigma=(\sigma^2/2\omega_0) \ln(\omega/\xi\omega_0)$. Then
\begin{align}
 &f(t_0)
\approx -\frac{\omega}{\omega_0} \left(\ln\frac{\omega}{e \xi \omega_0} 
- \frac{\sigma^2}{2\omega_0^2} \ln^2\frac{\omega}{\xi \omega_0}\right) \label{EqSPf0}\\
&f''(t_0)
 \approx -\omega \omega_0 \left(1+ \frac{\sigma^2}{\omega_0^2}\ln \frac{e\omega}{\xi \omega_0} \right)\notag.
\end{align}
Since energy transmitted to the bosons equals the MH gap, we insert $\omega=\Delta$. 
If we neglect also the contributions of order $\sigma^2/\omega_0^2$ then $\Gamma$ has especially compact form
\begin{equation} \label{EqSPGama}
\Gamma \approx 
g_{rc}^2 e^{-\xi}
\sqrt{\frac{2\pi}{\Delta\omega_0}}
\ \exp\left[
-\frac{\Delta}{\omega_0}
\ln\left(\frac{\Delta}{e \ \xi \omega_0}\right)
\right].
\end{equation} 
To test the applicability of Eq.~(\ref{EqSP0}) for our case we compare in Fig.~\ref{FigSPInt}: a) the numerical evaluation of $\Gamma$ from Eq.~(\ref{EqGammaInt}), b) the saddle point result for numerically (exactly) established saddle, c) the saddle point result for approximate saddle Eq.~(\ref{EqSPp}), and d) compact form of Eq.~(\ref{EqSPGama}) with $\sigma=0$.

\begin{figure}[ht]
\includegraphics[width=0.40\textwidth]{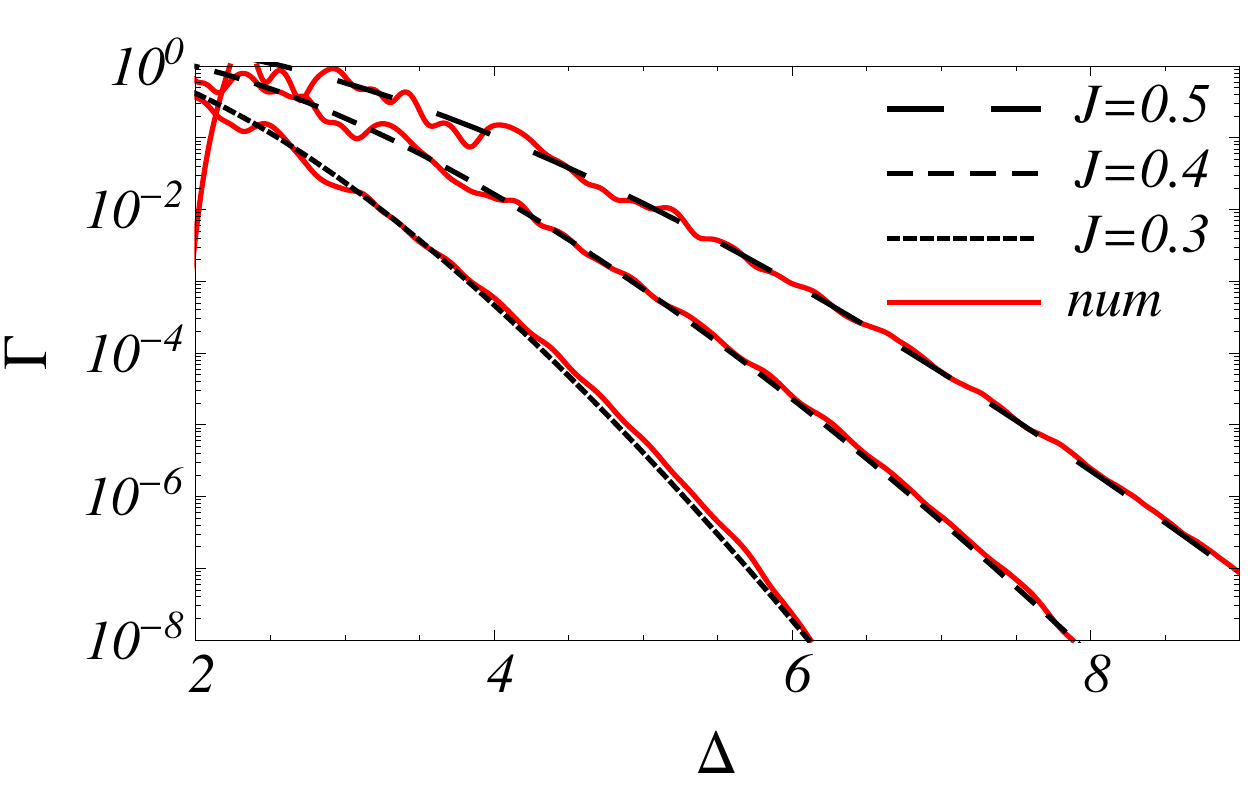}
\caption{(Color online) Fit of Eq.~(\ref{EqSPGama}) with $\xi,g_{rc}$ as the fitting parameters to the numerical result (num) for $\Gamma(\Delta)$ obtained on 2D system (as described in previous section) for $J=0.3,0.4,0.5$. }
\label{FigSPNum2D}
\end{figure}

\begin{figure}[ht]
\includegraphics[width=0.35\textwidth]{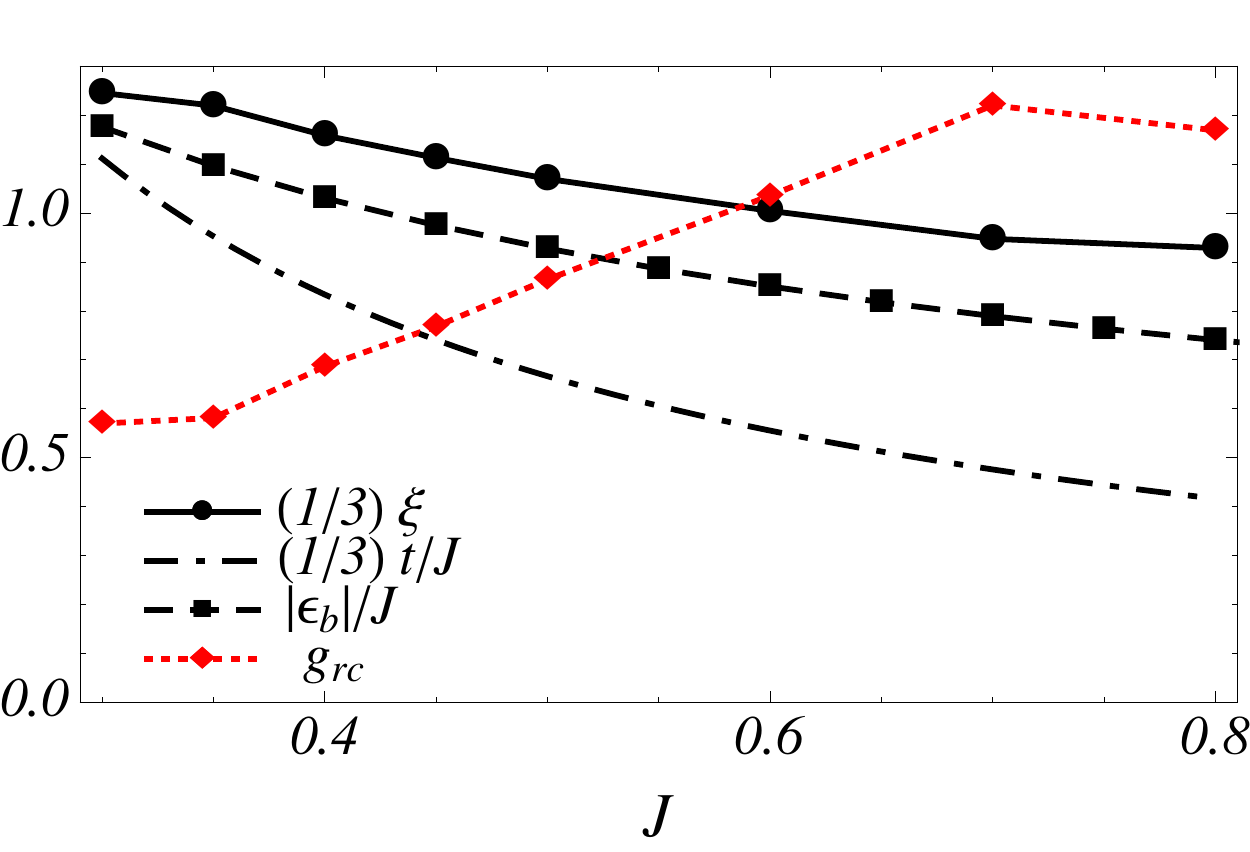}
\caption{(Color online) Values of the fitting parameters $\xi$ (boson coupling) and $g_{rc}$ (recombination prefactor) as a function of J. For comparison $|\epsilon_b|/J,t/J$ are plotted as well. Prefactor $1/3$ was used with $\xi$ and $t/J$ to unify the scales. }
\label{FigFitParam}
\end{figure}

Let us  apply Eq.~(\ref{EqSPGama}) to the HD exciton recombination due to the
emission of spin excitations studied in the previous sections. For that case we set $\omega_0=J$ and fit 
Eq.~(\ref{EqSPGama}) to the numerically obtained dependence $\Gamma(\Delta)$ for various $J$, 
with the dimensionless coupling $\xi$ and prefactor $g_{rc}$ as the fitting parameters. 
As shown in Fig.~\ref{FigSPNum2D}, formula Eq.~(\ref{EqSPGama}) captures the dependence $\Gamma(\Delta)$ for 
$\xi$ that is mildly dependent on $J$ (see Fig.~\ref{FigFitParam}).
This result has a fundamental importance since it signifies that the recombination of HD bound pair via multi-magnon
emission  can be described in a much broader frame - 
as a decay via many bosons. Besides the exponential form the 
most important message from Fig.~\ref{FigFitParam} is that the effective exciton-boson coupling is very strong $\xi \sim 3$.
The dependence of $\xi$ on $J$ resembles $\epsilon_b/J$,
where $\epsilon_b$ is numerically established binding energy of the HD pair, but with a substantially bigger prefactor. 
Latter relation is deduced from Eq.~(\ref{EqXi}) if we associate the HD pair binding energy with the exciton-boson binding energy, which might be oversimplified.
On the other hand, $\xi$ has milder $J$ dependence yet similar strength as $t/J$, which would emerge from the $n$-th order perturbation theory, Eq.~(\ref{fit2}), taking the charge-spin coupling to be simply the hopping term in Eq.~(\ref{EqHtj}). The prefactor dependence $g_{rc} \sim J$ is in qualitative agreement with the original model Eq.~(\ref{EqHrec}).

To give a definite comment on which approach, perturbation expansion or exciton-boson model, gives better description could be pointless since they must be essentially intertwined.
Still, consideration of exciton coupled to bosons elaborated in this section seems natural and the interpretation of the fitting parameters rather clear: $\xi$ can be identified as the exciton-boson coupling strength, whereas deviation of the value $\alpha_0\sim 0.8$, Eq.~(\ref{fit2}), from the expected $\alpha_0=2$ could not be argued properly \cite{lenarcic13}. However, probably both, discrepancy in $\alpha_0$ and lack of quantitative understanding of coupling $\xi$, originate in the nonperturbative nature of the charge-spin coupling.

\section{Comparison with experiments and discussion}

{\it Comparison with experiments:} 
When discussing the application of the theory to cuprates most parameters are well established. 
The $t$-$J$ model has been used by many authors for the quantitative comparison  of experimental 
results for various properties. In this sense quite well established parameters are $t\approx 0.35$~eV and $J/t \approx 0.4$, slightly varying within the cuprates. Since the MH gap (or more directly the optical gap) $\Delta_0$ is also determined by optical  
absorption, the only undetermined parameter is the prefactor $t_{rc}$, Eq.~(\ref{EqHrec}),
which we fix to the theoretically obtained $t_{rc}=J/2$. It should be noted that to get $\Delta$ relevant for the s-type bound state, as defined in Sec. \ref{SExcRecRate}, energy difference to the p-type unbound but optically active state has to be subtracted, 
$\Delta=\Delta_0-|\epsilon_b|$.
Most pump-probe results are so far obtained for two 2D undoped cuprates: NCO and LCO \cite{okamoto11}.
The characteristic microscopic unit time in these systems is given by
the elementary process of intercell hopping, i.e. $\tau_0= \hbar/t \sim 2$~fs.

\noindent NCO: Standard values quoted for NCO are \cite{okamoto11}:  the optical 
gap $\Delta_{0}=1.6~$eV and $J = 0.155$~eV,  
so that $\Delta=4.1~t$ and from Eq.~(\ref{EqGam}) $\Gamma  \sim 2.2 \cdot 10^{-2}/ \tau_0$.
Finally this leads to  $\tau=\Gamma^{-1} \sim 0.09~$ps, which is fairly close 
 to the experimentally measured $\tau \sim 0.2$~ps \cite{okamoto11}. 
 
\noindent LCO: Analogous values for LCO are: optical gap $\Delta_{0}=2$~eV and 
$J=0.133$~eV, so that $\Delta=5.3~t$ and $\Gamma\sim 1.3 \cdot 10^{-4}/\tau_0$,
yielding $\tau\sim 15$~ps. 
For this material detailed analysis was not performed, yet it is reported to have considerably longer relaxation \cite{okamoto11}, consistent with our result. From our theory the difference is quite evident, appearing due
to smaller $J$ and larger $\Delta_0$ in the case of LCO.

{\it Effective models: } 
The aim of our theoretical consideration of the problem is to establish the mechanism for the description of the recombination process of photoinduced charged particles in cuprates, based on a minimal sufficient model.
Rather than performing the calculations with the prototypical Hubbard model, we canonically transformed it, leading to the model defined by Eqs.~(\ref{EqHtj},\ref{EqHrec}). Its clear advantage is that
by separating sectors with different number of HD pairs in lowest order, as suggested by experimentally measured timescales of recombination,
a) it assists to extract the excitonic state of bound HD pair from the otherwise complex spectra of Hubbard model, b) takes into account that this state is not an eigenstate (and should therefore decay) in a transparent way - via the creation/recombination term, which serves as a perturbation causing the decay. Since undoped cuprates, being of primer interest of the whole discussion, are actually Mott insulators of the charge-transfer type we derived a similar minimal model also from a more realistic multi-band tight binding model including relevant Cu and O orbitals. Contrary to the previous studies of doped cuprates, hole- and electron-like excitations in this case have to be addressed on equal footing. As observed before the hole-electron (holon-doublon) symmetry is broken in such model \cite{feiner96}. However, the minimal model describing recombination has similar form with quantitatively comparable strength of operators causing decay of HD pairs as its single-band analogue. Only the internal structure of recombination/creation operators is somewhat reacher - allowing new intermediate states. From this we conclude that minimal model derived from the single-band Hubbard is sufficiently good, with a slight modification of Mott gap being interpreted as the charge-transfer gap.

{\it Existence of exciton:}
Our calculation of the recombination rate relies on the assumption that after being created holon and doublon form a s-type bound state on a timescale shorter than the recombination one. Besides observations in nonlinear optical susceptibility in LCO \cite{maeda04}, indirect experimental evidence for formation of such exciton is fluence (pump intensity) independent recombination rate with an exponential decay of charge density. If pairs were not bound, recombination process would depend on the probability to encounter the oppositely charged particle, evidently leading to a non-exponential decay. Since HD pair binds in order to minimize the distortion of short-range ordered spin background in its vicinity, the exciton should cease to exist in experimental conditions when the order is melted, e.g. when pumping the insulator with high fluence or well above the gap.

{\it Validity of Fermi golden rule:}
Usage of Fermi golden rule seems reasonable since recombination of charged particles is a slow process as compared to the scale $\hbar/t$ of the time-dependent simulations. Still, to test how important are the higher order terms that were neglected we performed the time-dependent evolution of initial excitonic state under Hamiltonian containing the recombination/creation term as well. We observe again an exponential decay of HD pair occupation number. On should beware that such calculation has its limitations too: a) discreteness of spectra sets upper bound for propagation due to recurrence of HD pair, b) virtual processes cause short time oscillations that destabilize the pair yet do not lead to true recombination, c) presence of perturbation alters the whole spectra, shifting the energies and leading to the reconsideration of the definition of the gap, d) we restricted the Hilbert space to the subspace of one and zero HD pairs. Still, the recombination rates obtained with both methods are comparable, and in the larger system with $N=26$ sites, where finite system artifacts are less pronounced, show slightly faster recombination in time-dependent calculation, as one would expect from the inclusion of additional processes.

{\it Origin of fast recombination:}
As a result of our study we can conclude that emission of spin excitations can be considered as a plausible
mechanism for the non-radiative recombination of photoinduced 
charges in a MH insulator, in spite of many bosons $n \sim \Delta/J \gg 1$
involved in a simultaneous emission.
Feasibility of creation of such large number of spin excitations itself has been demonstrated experimentally by the phonon assisted multimagnon light absorption \cite{perkins93,lorenzana99}.
The importance of analogous multi-phonon processes has been addresses theoretically as possibly relevant for decay in carbon nanotubes \cite{avouris06,perebeinos08}. However, the reason for much faster recombination in MH insulators as compared to the semiconductors \cite{yu99} is primarily in strong coupling between charged particles (holons and doublons) and spin background, in addition to obviously larger scale of spin excitations $J$ then the typical phonon energies $\omega_0$.
According to our understanding this strong coupling is manifested in two intertwined observations:
a) as revealed by the calculation of spin correlations already the HD exciton involves strong perturbation of the spin AFM background, which can be in the proces of recombination even further enlarged due to possible additional spin flips caused by $H_{rc}$,
b) on the level of effective exciton-boson Hamiltonian the relevant exciton-boson coupling turned out to be strong.

{\it Short-range vs. long-range order:}
It should be pointed out that the existence of the AFM long-range order and standard magnon excitations is not a necessary precondition for our analysis. The relevant excitations that receive the energy of HD pair are general multiple spin excitations or paramagnons, present also in the paramagnetic phase. All those excitations should have is dispersive nature in order to distribute the local spin perturbation. On the other hand, short-range spin correlations are necessary to provide the dressing of HD pair with spin excitations, and insure the existence of exciton. Other study\cite{sensarma11} of decay of unbound uncorrelated holon and doublon in completely spin disordered background revealed very slow recombination, proving the necessity of at least short-range correlated spin-background. After all, our calculations are done in small system which is big enough to accommodate the dressing of HD pair, however does not display long-range order in the strict sense. The role of latter is consequently not present in the result for recombination rate $\Gamma$.

{\it Higher photoexcited charge densities:}
Mechanism for recombination via emission of spin excitations should be relevant for systems with low density of photoexcited carriers that in such conditions presumably form HD excitons. In experiments using high fluence pump pulses, creating high density of photoexcited charge carriers, other mechanism might become dominant, e.g. so called Auger processes where energy of HD pair is transmitted to other charged carriers created within the pump. When sufficient density of charges is provided, dominance of such processes originates in easier instantaneous energy transmission - simply raising kinetic energy of remaining charge.
Clear experimental indication for such processes should be non-exponential decay of particle density, as long as what is observed is not only deviation around the thermal density of charges. The role of reversed, yet similar processes of holon-doublon pair ionization in the initial fast relaxation of doublons excited well above the gap has been established within the DMFT \cite{werner14}.
Moreover, related kinetic-assisted recombination mechanism, possibly consisting of several scattering processes, are dominant in experiments on fermionic cold atoms\cite{strohmaier10,sensarma10} and in DMFT studies\cite{eckstein11,eckstein12}.

{\it Role of dimension:}
In the present analysis the crucial ingredient for the fast recombination is strong charge-spin coupling.
This is inherently present within the 2D (also higher dimensional) strongly  correlated system, 
as modeled within the Hubbard model with $U\gg t$ or the $t$-$J$ model with $J<t$, where 
mobile photoexcited or doped charges crucially perturb and frustrate the spin background.   
On the other and, the physics in 1D correlated system 
could be quite different due to the phenomenon of charge-spin separation. It is established that 
e.g. within the 1D $t$-$J$ model the charge-spin coupling is quite ineffective and the motion
of holes/doublons is nearly free for $J \ll t$. Therefore other mechanisms, both for the exciton
formation as well as for the HD recombination, have to be invoked to deal with 
the photoexcited 1D MH insulators.

\begin{acknowledgements}
The authors acknowledge valuable discussions with T. Tohyama, R. McKenzie and D. Gole\v{z}.
This work has been supported by the Program P1-0044 and the project J1-4244 
of the Slovenian  Research Agency (ARRS). 
\end{acknowledgements}

\appendix{} 
\section{Intra-site Diagonalization for Charge-transfer Hubbard Model}\label{AppIntraStates}

Recombination/creation operator $H_{rc}$, Eq.~(\ref{EqRecCT}), derived from the original three-band Hamiltonian, Eq.~(\ref{EqHbare}), could have been obtained from higher order perturbative hopping processes, in a similar manner as the exchange coupling in Ref. \cite{muller-hartmann02}. Instead, our derivation of $H_{rc}$ is based on the introduction of states associated with a single cell, where each cell contains a Cu orbital and a Wannier O orbital. Those states represent holon and doublon as well as neutral states and are calculated as the eigenstates of single-cell Hamiltonian $H_{0i}$, Eq.~(\ref{EqHOneCell}). Coupling between cells is then established by the relevant matrix elements for states on adjacent cells, nontrivial due to hybridization between Cu and O orbitals in the single-cell states. The coupling strengths are set by the Hamiltonian Eq.~(\ref{EqHcc}) with Wannier-orbital transformation inherently present in the hopping parameters.
As originally proposed by \cite{feiner96}, the intra-cell diagonalization that gives us the single-cell states has to be performed within each total spin sector. 
In the doublet basis, Eq.~(\ref{EqDoub}),  we diagonalize the Hamiltonian
\begin{equation}
h^{1/2} =
\left( \begin{array}{cc}
0 & -\bar{t}_{pd}  \\
-\bar{t}_{pd} & \Delta_0 \\
\end{array} \right)
\end{equation}
yielding the g.s. $|g_{\sigma}\rangle$ that represents the charge-neutral (in the language of single-band Hubbard model spin-like) state with the energy $E_g$ 
\begin{align}
&|g_s\rangle=\cos\theta |\bar{d}_{s}\rangle + \sin\theta |p_s\rangle, \label{EqStDub}\\
&E_{g}=\frac{\Delta_0}{2}\left(1 - \sqrt{1+\tan^2(2\theta)}\right),
\end{align}
where
$\tan2\theta=2\bar{t}_{pd}/\Delta_0$.

\begin{figure}[ht]
\includegraphics[width=0.43\textwidth]{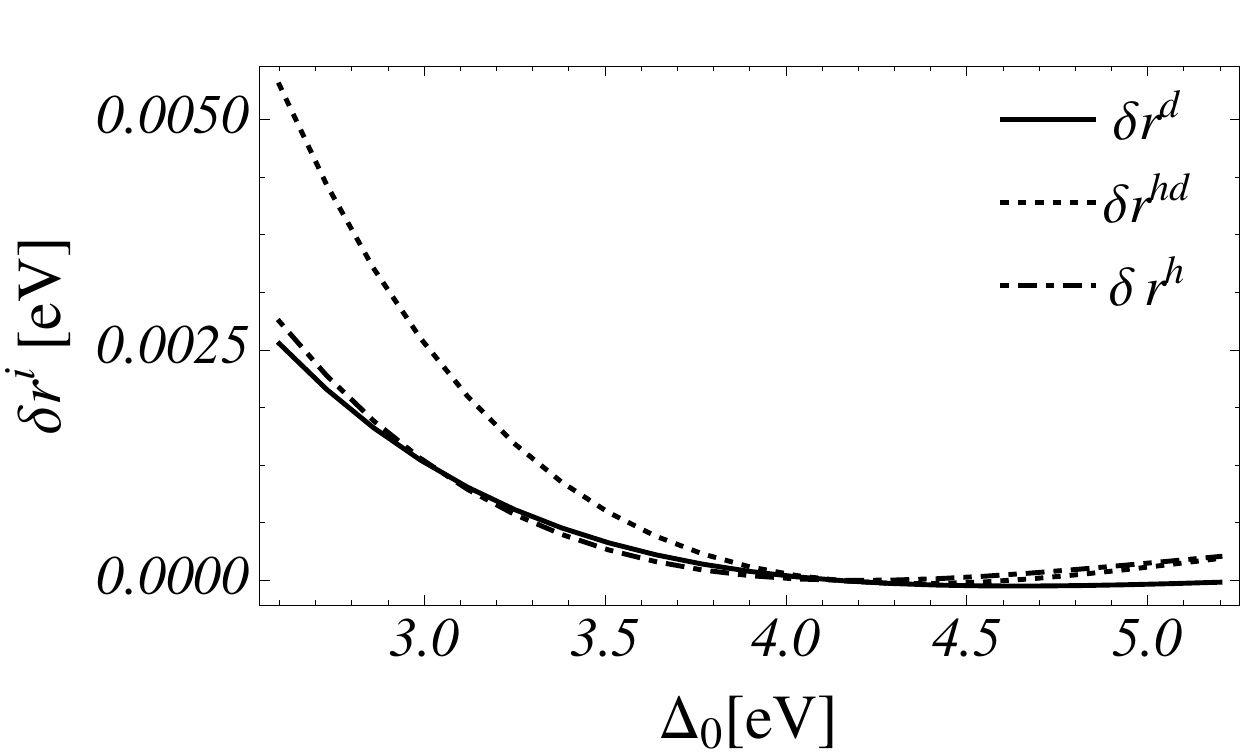}
\caption{The error in recombination coupling parameters, $\delta r=r_{num}-r_{appr}$, originating in approximate calculation Eq.~(\ref{EqStHol}) of holon state $|H\rangle$ and its energy $E_H$  as a function of $\Delta_0$. For other parameters standard values are used.}
\label{FigDrec}
\end{figure}

Within the singlet subspace, Eq.~(\ref{EqZR}),  holon is represented by the 
generalized Zhang-Rice singlet, which in addition to the dominant Zhang-Rice component
$(1/\sqrt{2})(\bar d_{\uparrow}^\dagger p_{\downarrow}^\dagger - \bar d_{\downarrow}^\dagger 
p_{\uparrow}^\dagger)|0\rangle$
contains also some fraction of
$\bar{d}_{\downarrow}^\dagger \bar{d}_{\uparrow}^\dagger|0\rangle, 
p_{\downarrow}^\dagger p_{\uparrow}^\dagger|0\rangle$
states. The fraction of each basis state is obtained by numerical diagonalization of the 
$3\times 3$ local Hamiltonian. 
Since $U_d\approx \bar{U}_p+2\Delta_0$ it turns out satisfactory to use basis
\begin{align}
\{|S_0\rangle &=\frac{1}{\sqrt{2}}(\bar{d}_{\uparrow}^\dagger p_{\downarrow}^\dagger - 
\bar{d}_{\downarrow}^\dagger p_{\uparrow}^\dagger)|0\rangle, \notag \\ 
|S_1\rangle & =\frac{1}{\sqrt{2}}(\bar{d}_{\uparrow}^\dagger \bar{d}_{\downarrow}^\dagger + 
p_{\uparrow}^\dagger p_{\downarrow}^\dagger)|0\rangle \}, 
\end{align}
in which local Hamiltonian is
\begin{equation}\label{EqHsinglet}
h^{0} =
\left( \begin{array}{cc}
\Delta_0 + \bar{V}_{pd} & -2\bar{t}_{pd} \\
-2\bar{t}_{pd} & \frac{1}{2}(U_{d} +\bar{U}_{p}) + \Delta_0. \\
\end{array} \right),
\end{equation}
yielding explicit expression for the holon state $|H\rangle$ and its energy 
\begin{align}
|H\rangle&=\cos\phi \ |S_0\rangle + \sin\phi \ |S_1\rangle, \label{EqStHol}\\ 
E_{H}&=\Delta_0 + \bar{V}_{dp}+\frac{U_d+\bar{U}_p-2\bar{V}_{pd}}{4}\left(1 - \sqrt{1+\tan^2 (2\phi)}\right),\notag
\end{align}
where 
$\tan2\phi=8\bar{t}_{pd}/(U_d+\bar{U}_p-2\bar{V}_{pd})$.
In order to check how much such approximation effects the recombination couplings Eq.~(\ref{EqRd}) for different channels, we compared those values if $|H\rangle$ and $E_H$ are calculated accurately by numerical diagonalization of $3\times 3$ Hamiltonian, or within the latter approximation. The difference in coupling strengths $\delta r=r_{num}-r_{appr}$ is not substantial, as shown in Fig.~\ref{FigDrec}. 

The triplet states $|T_s\rangle$ within each cell are decoupled and have energy $E_T=\Delta_0+\bar{V}_{pd}$.

\section{Effective Hopping Parameters for Charge-transfer Hubbard Model}\label{App3BandParams}
Hopping parameters that are introduced in the reduced single-band-like Hamiltonian, Eqs.~(\ref{EqHt3},\ref{EqHrc3}), are obtained by evaluation of matrix elements for the inter-cell Hamiltonian $H_{cc}$, Eq.~(\ref{EqHcc}), between the states $|H\rangle, |D\rangle,|T_s\rangle,|g_{s}\rangle $, Eqs.~(\ref{EqStHol},\ref{EqStTrip},\ref{EqStDub}), on adjacent sites. 
For example, parameter $t^h$ associated with hopping of holon is calculated from the matrix element 
$\langle H_{i}, g_{js}|H_{cc}|g_{is},H_j\rangle$.
Parametrized by $\theta,\phi$ and $\tilde\tau=2 t_{pd}\mu_{01}, \tau'=2 t_{pp}\nu_{01}$ they are presented in the Table \ref{TblHoppPar}.
\newline
 \begin{table}[htbp]
 \begin{center}
{\renewcommand{\arraystretch}{1.4}
\begin{tabular} {|l|l|}
\hline	Holon hopping	&	$t^{h}=t^{h}_d + t^h_p;$\\[2pt]
										&$t^{h}_{d}=\tilde\tau(\sin 2\theta + \sin 2\phi)/2$, \\ 
										& $t^{h}_{p}=\tau'\cos^2(\theta-\phi)/2.$\\ 
\hline	Doublon hopping	&	 $t^d=t^d_d + t^d_p;$\\[2pt]
										&$t^d_d=\tilde\tau\sin2\theta, \quad t^{d}_{p}=\tau' \sin^2\theta.$\\
\hline	Triplet hopping		&$t^{T_0}=t^{T_0}_d + t^{T_0}_p, \quad t^{T_1}=t^{T_1}_d + t^{T_1}_p;$\\ [2pt]
										&$t^{T_0}_d=\tilde\tau \cos 2\theta \sin\phi/2$, \\
										& $t^{T_1}_d=\tilde\tau\cos 2\theta \sin \phi/\sqrt{2}.$\\
										&$t^{T_{0}}_p=\tau'\cos\theta \cos(\theta-\phi)/2,$ \\
										& $t^{T_{1}}_p=\tau'\cos\theta \cos(\theta-\phi)/\sqrt{2}.$\\
\hline	Holon-doublon  & $t^r=t^r_d + t^r_p;$  \\[2pt]
				recombination& $t^r_d=\tilde\tau(\cos\phi + \sin2\theta \sin\phi)/\sqrt{2}$,\\
										& $t^{r}_{p}=\tau'\cos(\theta-\phi)\sin\theta/\sqrt{2}.$\\
\hline  Triplet-doublon  &  $t^{r_0}=t^{r_0}_d+t^{r_0}_p, \quad t^{r_1}=t^{r_1}_d+t^{r_1}_p$; \\ [2pt]
			recombination	& $t^{r_0}_d=\tilde\tau\cos2\theta/\sqrt{2}$,\\ 
										& $t^{r_1}_d=\tilde\tau\cos 2\theta,$ \\
										& $t^{r_0}_p=\tau'\sin2\theta/2\sqrt{2}$,\\
										& $t^{r_1}_p=\tau'\sin2\theta/2.$\\
\hline
\end{tabular} 
}
\end{center}
\caption{\label{TblHoppPar}Hopping parameters for reduced single-band-like Hamiltonian, Eqs.~(\ref{EqHt3},\ref{EqHrc3}), parametrized by $\theta,\phi$ and $\tilde\tau=2 t_{pd}\mu_{01}, \tau'=2 t_{pp}\nu_{01}$. }
\end{table}

\vspace{0.1cm}
These effective hopping parameters are together with the relative energies $\epsilon_H=E_H-E_g,\epsilon_D=-E_g,\epsilon_T=E_T-E_g$
the essential ingredient of recombination coupling strengths, as explicitly written in Eqs.~(\ref{EqRd}).

\clearpage

\vskip 5truecm
\bibliographystyle{physrev4}

\end{document}